\documentclass[usenatbib]{mn2e}

\usepackage{amsmath}
\usepackage{natbib}
\usepackage{epsfig}
\usepackage{txfonts}
\usepackage{subfigure}

\newcommand{\id}{{\,\rm d}}

\newcommand{\beq}{\begin{equation}}   %

\newcommand{\eeq}{\end{equation}}   %

\newcommand{\beqa}{\begin{eqnarray}}   %

\newcommand{\eeqa}{\end{eqnarray}}   %

\newcommand{\beal}{\begin{align}}

\newcommand{\enal}{\end{align}}

\newcommand{\bspl}{\begin{split}}

\newcommand{\espl}{\end{split}}

\newcommand{\bsub}{\begin{subequations}}

\newcommand{\esub}{\end{subequations}}

\newcommand{\bmulti}{\begin{multline}}   %

\newcommand{\beqm}{\begin{mathletters}}   %

\newcommand{\eeqm}{\end{mathletters}}   %

\newcommand{\me}{m_{\rm e}}

\newcommand{\Te}{T_{\rm e}}

\newcommand{\Tg}{T_{\gamma}}

\newcommand{\pd}{\partial}

\newcommand{\pAb}[2]{\frac{\displaystyle\pd #1}{\displaystyle\pd #2}}
\newcommand{\pAbc}[3]{\left.\frac{\displaystyle\pd #1}{\displaystyle\pd #2}\right|_{#3}}

\newcommand{\PAb}[3]{\frac{\displaystyle\pd^{#3} #1}{\displaystyle\pd {#2}^{#3}}}

\newcommand{\Abl}[2]{\frac{{\rm d} #1}{{\rm d} #2}}

\newcommand{\pot}[2]{#1 \times 10^{#2}}

\newcommand{\ion}[2]{{\text{{\sc #1}\,{\sc #2}}}}

\newcommand{\nbb}{{n^{\rm pl}}}

\newcommand{\twosav}[1]{{\left<\,#1\,\right>^{2\gamma}_{\rm 2s}}}

\newcommand{\twosavg}[3]{{\left<\,#1\,\right>^{#2}_{#3}}}

\usepackage{color}
\newcommand{\changeI}[1]{{#1}}

\title[Towards a complete treatment of recombination]
{Towards a \changeI{complete} treatment of the cosmological recombination problem}

\author[Chluba and Thomas]{J. Chluba$^{1,2}$\thanks{E-mail:
   jchluba@cita.utoronto.ca} and R.~M.~Thomas$^{1}$\thanks{E-mail:
  thomas@cita.utoronto.ca}
  \\
$^{1}$ Canadian Institute for Theoretical Astrophysics, 60 St. George Street,
Toronto, ON M5S 3H8, Canada\\
$^{2}$ Max-Planck Institut f\"ur Astrophysik, Karl-Schwarzschild-Str. 1,
D-85740 Garching, Germany}

\begin{document}

\date{Received 2010 October 18; Accepted 2010 October 27}

\maketitle

\begin{abstract}
  A new approach to the cosmological recombination problem is
  presented, which completes our previous analysis on the effects of
  two-photon processes during the epoch of cosmological hydrogen 
  recombination, accounting for $n$s-1s and $n$d-1s Raman events and
  two-photon transitions from levels with $n\geq2$.
  The recombination problem for hydrogen is described using an {\it
  effective} 400-shell multi-level approach, to which we subsequently
    add {\it all} important recombination corrections discussed in the
    literature thus far.
    We explicitly solve the radiative transfer equation of the
    Lyman-series photon field to obtain the required modifications to
    the rate equations of the resolved levels.
    In agreement with earlier computations we find that 2s-1s Raman
    scattering leads to a delay in recombination by $\Delta N_{\rm
      e}/N_{\rm e}\sim 0.9\%$ at $z\sim 920$.
Two-photon decay and Raman scattering from higher levels ($n>3$)
result \changeI{in small} additional modifications, and precise results can be
obtained when including their effect for the first $3-5$ shells.
This work is a major step towards a new cosmological
recombination code \changeI{({\sc CosmoRec})} that supersedes the physical model included in {\sc
  Recfast}, and which, owing to its short runtime, can be used in the
analysis of future CMB data from the {\sc Planck} Surveyor.

\end{abstract}


\section{Introduction}
\label{sec:Intro}
The {\sc Planck} Surveyor\footnote{www.rssd.esa.int/Planck} is
currently observing the temperature and polarization anisotropies of
the cosmic microwave background (CMB), and scientists all over the
world eagerly await its first data release, scheduled for early 2011.
With \changeI{{\sc Planck}} data sets cosmologists will be able to determine  key
cosmological parameters with unprecedented precision, making it possible to
distinguish between the {various models} of {\it inflation}
\citep[e.g. see][for recent constraints]{Komatsu2010}.

Over the past 5 years, various groups
\citep[e.g. see][]{Dubrovich2005, Chluba2006, Kholu2006, Switzer2007I,
  Wong2007, Jose2008, Karshenboim2008, Hirata2008, Chluba2008a,
  Jentschura2009, Labzowsky2009, Grin2009, Yacine2010} have realized
that the uncertainties in the theoretical treatment of the
cosmological recombination process could have important consequences
\changeI{for} the analysis of the CMB data from the {\sc Planck} Surveyor.
It was shown that in particular our ability to measure the precise
value of the spectral index of scalar perturbations, $n_{\rm S}$, and
the baryon content of our Universe will be compromised if
modifications to the recombination model by {\sc Recfast}
\citep{SeagerRecfast1999, Seager2000} are neglected \citep{Jose2010}.

To ensure that uncertainties in the cosmological recombination model
do not undermine the science return of the {\sc Planck} satellite,
it is crucial to incorporate all important processes leading to
changes in the free electron fraction close to the maxima of the
Thomson visibility function \citep{Sunyaev1970} by more than $\sim
0.1\%$ into {\it one} recombination module.
The main obstacle towards accomplishing this so far was that detailed
recombination calculations took too long to allow accounting for
the full cosmological dependence of the recombination corrections on a
model-by-model basis.
This led to the introduction of {\it improved fudge factors} to
{\sc Recfast} \citep{Wong2007, Wong2008}, or {\it multi-dimensional
  interpolation schemes} \citep{Fendt2009, Jose2010}, that allow fast
but approximate representation of the full recombination code.

Although it was already \changeI{argued} that for the stringent error-bars of today's
cosmological parameters such approaches should be sufficient
\citep{Jose2010}, from a physical stand point it would be much more
satisfying to have a full representation of the recombination problem,
that does not suffer from the limitations mentioned above, while
capturing all the important physical processes simultaneously.
\changeI{Furthermore, such a recombination module increases the flexibility, 
and allows us to provide extensions, e.g., to account for the effect of {\it dark matter annihilation}, energy injection by {\it decaying particles} \citep[e.g. see][]{Chen2004, Padmanabhan2005, Chluba2010a}, or the {\it variation of fundamental constants} \citep{Kaplinghat1999, Galli2009b, Scoccola2009}, while treating all processes simultaneously.}

In this paper, we describe our \changeI{new} approach to the recombination problem,
which enables us to fulfill this ambition by overcoming the problems
mentioned above.
Our code, \changeI{called {\sc CosmoRec}\footnote{\changeI{This code will be available at www.Chluba.de/CosmoRec.}}}, \changeI{runs in} $1-2$ minutes for a given set of
cosmological parameters as it stands and can be optimized further to
run well below a minute, eliminating the need for {\it fudge factors}
to solve the recombination problem.  One of the key ingredients that
facilitates this increase in speed is the effective multi-level
approach, which was proposed recently by \citet{Yacine2010}.

We also extend our previous analysis on the effects of two-photon processes
during the cosmological recombination epoch of hydrogen \changeI{\citep{Chluba2008a, Chluba2009}} to account for
$n$s-1s and $n$d-1s Raman scattering and two-photon transitions from
levels with $n\geq2$.
The radiative transfer equation for the Lyman-series photons during
hydrogen recombination is solved in detail using a PDE solver that we
developed for this purpose \changeI{and} can accommodate non-uniform grids \changeI{(see Appendix~\ref{app:PDE_solver} for more details)}.
Our results for the effect of Raman scattering on the recombination
dynamics are in good agreement with earlier computations
\citep{Hirata2008}.
Furthermore, we show that two-photon decays from levels with
$n\gtrsim 4-5$ can be neglected and Raman scattering is only important
for the first few shells.  

\changeI{The main difficulty with two-photon and Raman processes during the recombination epoch is the presence of resonances in the interaction cross-sections related to normal '1+1' photon transitions that are already included into the multi-level recombination code \citep{Switzer2007III, Chluba2008a, Hirata2008, Chluba2009}.
Unlike for the 2s-1s two-photon decay, all the higher $n$s-1s and $n$d-1s two-photon channels include '$1+1$' photon sequences via energetically {\it lower} Lyman-series resonances, i.e., $n{\rm s/d} \leftrightarrow k{\rm p}\leftrightarrow {\rm 1s}$ with $k<n$.
Similarly, for $n$s-1s and $n$d-1s Raman-scattering events {\it all higher} Lyman-series resonances, i.e., $n{\rm s/d} \leftrightarrow k{\rm p}\leftrightarrow {\rm 1s}$ with $k>n$, appear.
Therefore, special care has to be taken to avoid {\it double-counting} of these resonances in the rate equations of the multi-level atom, as we explain in \S~\ref{sec:double_1}, \S~\ref{sec:double_2}, \S~\ref{sec:double_3}, and \S~\ref{sec:double_4}.}

In \S~\ref{sec:pert_approach} we outline our principle approach to the 
recombination problem.
The terms for the radiative transfer equation that allow to take all
important recombination corrections into account are derived in
\S~\ref{sec:DNnu_pert}.
We then solve the evolution equation for the high frequency photon
field during the recombination epoch, and illustrate the different
changes in \S~\ref{sec:DI-Ly-n_cases}. In \S~\ref{sec:DNe_cases}, we
discuss the different corrections to the ionization history, and we
present our conclusions and outlook in \S~\ref{sec:conc}.

\section{Perturbative approach to solving the full recombination problem}
\label{sec:pert_approach}

\subsection{General aspects of the standard recombination problem}
\label{sec:general}
\label{sec:rate_equations}
{The cosmological recombination problem consists of determining an accurate
estimate of the free electron fraction, $X_{\rm e}=N_{\rm e}/N_{\rm H}$, as a function of
redshift.} Because of particle conservation, and the number of
electrons in excited states of \ion{H}{i} and \ion{He}{i} being
negligible, one {may} write\footnote{We {assume to be} at redshifts $z \ll 6000$,
  well after doubly ionized helium recombines.}
\beal
\label{eq:Sol_Ne}
N_{\rm e}\approx N_{\rm H}[1-X^{\rm H}_{\rm 1s}]+N_{\rm H}[f_{\rm He}-X^{\rm He}_{\rm 1s}],
\end{align}
where $N_{\rm H}$ denotes the number density of hydrogen nuclei, and
$f_{\rm He}=N_{\rm He}/N_{\rm H}$ is the fraction of helium
nuclei. The populations of the different levels are given by $X^{\rm
  a}_i=N^{\rm a}_i/N_{\rm H}$, where 'a' indicates the atomic species.
Furthermore, $X_i\equiv X^{\rm H}_i$ for convenience.

Equation ~\eqref{eq:Sol_Ne} implies that the recombination problem
reduces to finding solutions to $X^{\rm a}_{\rm 1s}$.
For hydrogen, the {\it standard} rate equation describing the
evolution of the ground-state population has the form \citep[see
  also;][]{SeagerRecfast1999, Seager2000}
\bsub
\label{eq:pert_X1s_stand}
\beal
\label{eq:pert_X1s_stand_a}
\left. \Abl{X^{\rm H}_{\rm 1s}}{t} \right|_{\rm st} 
&= \Delta R^{\rm st}_{\rm 2s \leftrightarrow 1s} + \sum_k \Delta R^{\rm st}_{k\rm p \leftrightarrow 1s}
\\
\label{eq:pert_X1s_stand_b}
\Delta R^{\rm st}_{\rm 2s \leftrightarrow 1s}&=A_{\rm 2s1s} \left[ X^{\rm H}_{\rm 2s} - X^{\rm H}_{\rm 1s} e^{-h\nu_{21}/k\Tg} \right]
\\[2mm]
\label{eq:pert_X1s_stand_c}
\Delta R^{\rm st}_{k\rm p \leftrightarrow 1s}&=A_{k\rm p1s}   (1+n^{\rm pl}_{k\rm p 1s})
 \left[ X^{\rm H}_{k\rm p} - 3\,X^{\rm H}_{\rm 1s} \bar{n}_{k \rm p 1s} \right].
\end{align}
\esub
Here $\bar{n}_{k\rm p 1s}$ is the mean photon occupation number over
the Lyman-$k$ line profile, $A_i$ the atomic rate coefficients for
spontaneous emission, and $n_{k\rm p 1s}^{\rm pl}$ is the occupation
number of the {CMB} blackbody photons at
the Lyman-$k$ transition frequency $\nu_{k 1}\equiv \nu_{k\rm p 1s}$.

The solution of Eq.~\eqref{eq:pert_X1s_stand}, depends on the level
populations of the 2s- and $k$p-states. \changeI{In addition,} the photon
distribution in the vicinity of every Lyman-resonance \changeI{has to be known}, to define
$\bar{n}_{k\rm p 1s}$. \changeI{$\bar{n}_{k\rm p 1s}$} is often estimated {by} the Sobolev
approximation, \changeI{which, however, breaks} down during
recombination, leading to non-negligible corrections to the
recombination dynamics \citep[e.g. see][]{Chluba2008b, Chluba2009,
  Chluba2009b}. The rate equations for the 2s- and $k$p-states
themselves can, in principle, be explicitly {given}. But here it is {only}
important to realize that {these} lead to a large network of rate
equations which depends on the populations of {\it all other} excited
levels. To complicate matters further the electron temperature, $\Te$,
enters the whole problem via recombination coefficients,
$\alpha_{i}(\Te,\Tg)$, to each level $i${, where} $\Tg$ is the photon
temperature.

\changeI{The evolution of $\Te$ is described by one simple differential equation, which accounts for the cooling of electrons caused by the Hubble-expansion, and the energy exchange with CMB photons via Compton scattering. Other processes, e.g., such as Bremsstrahlung cooling, are subdominant \citep{Seager2000}.}

\subsubsection{The effective multi-level approach}
\label{sec:Yacine}
%
{Recently,} \citet{Yacine2010} suggested to simplify the recombination problem
to a subset of levels that need to be followed explicitly. 
{Here} we shall call the members of this subset 'resolved' levels.
This approach enables us to account for the effect of recombinations
due to highly excited states ($n>100$), without actually solving for
all these level populations explicitly. 
The rationale being that except for the optically thick Lyman-series transitions, all other
rates are mediated by the CMB blackbody photons, and hence only depend
on the photon and electron temperatures.

The downside of this simplification entails the need to tabulate
{\it effective} rate coefficients as a function of $\Tg$ and $\Te$
prior to the computation. This however needs to be done only once, and
given that the number of resolved states necessary for converging
solutions is small, this {\it effective multi-level approach} results
in tremendous {speed-ups} for recombination calculations
  \citep[see][for more details]{Yacine2010}.
For this work we {also} implemented such an effective rate approach.
The rate coefficients for an effective 400-shell hydrogen atom were
computed using our most recent recombination code \citep{Chluba2010}{, while helium} is described with a multi-level treatment \citep{Chluba2009c}.

Within this framework the choice of the number of resolved states
depends on the extra {\it physics} that one intends to include.
For example, in a {\it minimal} model for the hydrogen recombination
problem one {should} explicitly solve for the\footnote{Because the 2s and
  2p states are usually close to full statistical equilibrium, one
  could also eliminate either of {these} states using $X_{\rm 2p}=3 X_{\rm
    2s}$ and, as a result, closely resemble the normal {\sc Recfast}
  code, now without requiring a fudge factor.} 1s, 2s, and 2p
level populations in tandem with the electron temperature, $\Te$.

This minimal choice already allows us to include processes that affect
the net rates in the 2s-1s two-photon channel and the 2p-1s
Lyman-$\alpha$ resonance, \changeI{e.g.,} the effect of stimulated 2s-1s
two-photon decay \citep{Chluba2006}, or the feedback of Lyman-$\alpha$
photons on the 1s-2s rate \citep{Kholu2006}.
However, since we restricted ourselves to the 1s, 2s and 2p states,
corrections due to Lyman-$\beta$ or higher resonance feedback cannot
be modelled as these would require resolving $n$p states with $n>2$
\citep{Chluba2007b}. We will return to these points in
\S~\ref{sec:method}.

\subsubsection{Sobolev approximation for $\bar{n}_{k\rm p 1s}$}
\label{sec:rate_eq_X1s}
In a multi-level approach the Sobolev approximation
is invoked to obtain a solution for the photon-field around every
resonance appearing in Eq.~\eqref{eq:pert_X1s_stand_c}. The photon
occupation number around each line is {then} given by\footnote{We assumed
  that as $\nu\rightarrow \infty$ the distortion $\Delta n_\infty
  \rightarrow 0$, and that {(as usual)} the factor $(\nu_{k1}/\nu)^2\sim 1$.}
\beal
\label{eq:Dnnu_Sobolev}
\Delta n^k_\nu =\Delta n^k_{\rm L} \left[ 1- e^{-\tau^k_{\rm S} [1-\chi^k_\nu]} \right],
\end{align}
where $\Delta n^k_{\rm L}\approx n^k_{\rm L} -n^{\rm pl}_{{k\rm p
    1s}}$, $\tau^k_{\rm S}$ is the Sobolev optical depth in the
Lyman-$k$ resonance, and $\chi^k_{\nu}=\int_0^\nu \varphi^k_{\rm
  V}(\nu)\id \nu$. Here $\varphi^k_{\rm V}(\nu)$ is the Voigt profile
corresponding to a resonance, and the line occupation number, $n^k_{\rm
  L}$, is defined as:
\beal
\label{eq:Dn_L}
n^k_{\rm L} 
=\frac{N_{k \rm p}}{3\,N_{\rm 1s}}.
\end{align}
Consequently, a simple approximation for the mean occupation number is
\beal
\label{eq:nbar_i}
\bar{n}_{k\rm p1s} =n^k_{\rm L} - P^k_{\rm S} \Delta n^k_{\rm L},
\end{align}
with $P^k_{\rm S}=[1-e^{-\tau^k_{\rm S}}]/\tau^k_{\rm S}$ being the
Sobolev escape probability.

For the Lyman-$\alpha$ resonance Eq.~\eqref{eq:Dnnu_Sobolev} results in {a} photon distribution that is
rather unphysical \citep[e.g. see discussion in][]{Chluba2008b}. This
is primarily due to the assumption that every interaction with the
resonance leads to a complete redistribution of photons over the whole
line profile, which for typical values of $\tau_{\rm S}$ during
recombination couples the photon distribution from the line center up
to frequencies in the Lyman-continuum.
For conditions present in our {Universe,} photon redistribution {over frequency} is
much less effective, most notably in the distant wings. Thus, it is
important to distinguish between scattering, real emission and
absorption events, as we will discuss in \S~\ref{sec:DNnu_pert}.

\subsection{Beyond the standard rate equation for 1s}
As mentioned in \S~\ref{sec:Yacine}, within the effective
multi-level approach the choice for the resolved states depends on the
physics to be modelled in detail.
For example, in order to include the full effect of Lyman-series
feedback, say up to $n=8$, the {\it minimal} model that follows 1s,
2s, and 2p would {at least} have to be extended by all $k$p-states up to 8p.

Also, the inclusion of two-photon processes from higher levels and
Raman-scattering, requires us to re-write
equation~\eqref{eq:pert_X1s_stand} in a more generalized form as,

\beal
\label{eq:pert_X1s_general}
\left. \Abl{X^{\rm H}_{\rm 1s}}{t} \right|_{\rm mod}
&= \sum_i \left\{ X^{\rm H}_i R_{i\rightarrow \rm 1s} - X^{\rm H}_{\rm 1s}  R_{{\rm 1s} \rightarrow i} \right\},
\end{align}
where $R_{i\rightarrow j}$ are the rates between the levels $i$ and
$j$. These rates depend on atomic physics, the CMB blackbody, the
electron temperature, and the solution for the Lyman-series spectral
distortion introduced by the recombination process.

To include two-photon corrections to the Lyman-series up to $n\leq
n_{\rm max}$, the important levels to follow are all the $n$d and
$n$s-states {with $ 2 \leq n \leq n_{\rm max}$}. The corresponding
partial rates to the $n$p-states drop out of the equations, and the
Lyman-series emission and absorption profiles, usually given by a Voigt
{function}, will be replaced by the two-photon profiles for the $n{\rm
  s}\leftrightarrow{\rm 1s}$ and $n{\rm d}\leftrightarrow{\rm 1s}$
{processes,} and similarly for the Raman process. We will specify
these corrections more precisely in the following sections.

\subsubsection{Accounting for corrections from radiative transfer effects}
\label{sec:method}
Changes in the level populations, electron temperature and free
electron fraction remain small ($\sim 1\%$), when different physical
processes, which were neglected in earlier treatments
\citep[e.g. see][for overview]{Jose2010} are included. This justifies
treating corrections to $\Te$ and the populations of resolved levels,
$X_{i}$, as small {\it perturbations}.
On the other hand, it has been shown that the changes in
the photon field caused by time-dependence \citep{Chluba2008b}, line
scattering \citep{Chluba2009b, Hirata2009}, or two-photon corrections
\citep{Hirata2008, Chluba2009}, are {\it non-perturbative}.

In \S\ref{sec:DNnu_pert} we derive in detail the different
  correction terms for the photon diffusion equation and provide
  modifications to the net rates of the effective multi-level atom.
The idea is to first solve the recombination history using the
effective multi-level approach in the '$1+1$' photon
  description, i.e. equate $R_{i\rightarrow j}=R_{i\rightarrow
    j}^{1+1}$ in Eq.~\eqref{eq:pert_X1s_general}, and then compute the
  solution to the photon field using the radiative transfer
  equation. 
This then leads to corrections in the net rates, {which are} used in computing
changes to the recombination dynamics, and hence {modify}
Eq.~\eqref{eq:pert_X1s_general}.
These corrections being small, demand only one iteration to
converge. 
Detailed descriptions to the notations in the following sections and
part of the methods used can be also found in \citet{Chluba2008b,Chluba2009,Chluba2009b}.

\section{Equation for the photon field evolution and corrections to the effective multi-level atom}
\label{sec:DNnu_pert}
To account for all corrections to the cosmological recombination
problem, it is important to follow the evolution of {\it non-thermal
photons} in the Lyman-series, which are produced during recombination.
These photons interact strongly with neutral hydrogen atoms throughout
the entire epoch of recombination, and their rate of {\it escape} from
the Lyman-resonances is one of the key ingredients in accurately
solving the recombination problem.

The partial differential equation governing the evolution of the
photon field has the form \citep[see][for a detailed
  discussion]{Chluba2008b}
\beal
\label{eq:pert_Dnnu}
 \frac{1}{c}\left[\pAbc{\Delta n_\nu}{t}{\nu} - H\nu \pAbc{\Delta n_\nu}{\nu}{t}\right]
 &=
 \mathcal{C}[\Delta n_\nu]_{\rm em/abs} + \mathcal{C}[\Delta n_\nu]_{\rm scat},
 \end{align}
where $\Delta n_\nu=\frac{c^2}{2\nu^2}\Delta N_\nu$ is the distortion
in the photon occupation number, and a distinction is made between the
collision terms leading to emission and absorption,
$\mathcal{C}[\Delta n_\nu]_{\rm em/abs}$, and scattering,
$\mathcal{C}[\Delta n_\nu]_{\rm scat}$. 
As an example, the first term on the right hand side of the equation can account {for} two-photon
corrections to the line {profiles,} while the second term, electron and/or
resonance scattering.
The second term on the left hand side describes the redshifting of
photons due to Hubble expansion, and plays a crucial role in the
escape of photons from the optically thick Lyman-series resonances.

In Eq.~\eqref{eq:pert_Dnnu} the CMB blackbody has been subtracted,
i.e., $\Delta n_\nu=n_\nu-n^{\rm pl}_\nu$, where $n^{\rm
  pl}_\nu=[e^{h\nu/k\Tg}-1]^{-1}$ is the blackbody occupation {number,}
because the left hand side directly vanishes for a blackbody with
temperature $\Tg(z)\propto (1+z)$. Also, spectral distortions created
by Compton scattering off electrons with $\Te \neq \Tg$ will be
extremely small for conditions in our Universe\footnote{The cooling of
  CMB photons by losing energy to keep electrons at $\Te \sim \Tg$
  should lead to a $y$-distortion with $y$-parameter $y\sim
  10^{-10}-10^{-9}$. The dissipation of energy by acoustic {waves} should
  lead to $y\sim 10^{-8}$. These can be neglected for our purpose.},
at least if there is no additional energy release.

By changing the time-variable to redshift {$z$ using} $\id z/ \id t=-
H(1+z)$, and scaling to dimensionless frequency $x=\nu/\nu_{\rm 21}$,
Eq.~\eqref{eq:pert_Dnnu} reads
\beal
\label{eq:pert_Dnnu_z}
\pAbc{\Delta n_x}{z}{x}\!=\! - \frac{x}{[1+z]} \pAbc{\Delta n_x}{x}{z} 
- \Lambda_z\left\{ \mathcal{C}[\Delta n_\nu]_{\rm em/abs} + \mathcal{C}[\Delta n_\nu]_{\rm scat} \right\},
 \end{align}
where $\Delta n_x= \nu_{\rm 21}\,\Delta n_\nu$ and $\Lambda_z=\frac{c
  \nu_{\rm 21}}{H [1+z]} $. We will now discuss the terms describing
the resonance and electron scattering.  In \S~\ref{sec:DNe_sc} we specify
the different emission and absorption terms, which then in
\S~\ref{sec:DI-Ly-n_cases} and \S\ref{sec:DNe_cases} are used 
to compute the corrections to the Lyman-series distortion and
ionization history.

\subsection{Inclusion of partial redistribution by resonance and electron scattering}
\label{sec:DNe_sc}
Here we provide the terms for the Boltzmann equation describing the
effect of (partial) photon redistribution by resonance and electron
scattering.
The form of the collision term for these cases within a Fokker-Planck
formulation was discussed earlier \citep[e.g.][]{Zeldovich1969,
  Basko1978a, Basko1978b, RybickiDell94, Sazonov2000, Rybicki2006,
  Chluba2009b}.
Since we are only following the evolution of the distortion from a
blackbody, and since it is clear that induced effects are
negligible\footnote{Eliminating the dominant term of the CMB blackbody
  leaves us with a term that is tiny because we are always in
  the distant Wien tail of the CMB at all times during
  recombination.}, one can readily write
\beal
\label{app:coll_n_nu_scatt}
\Lambda_z \!\left.\mathcal{C}[\Delta n_{\nu}]\right|_{\rm scatt}
&\approx
\frac{1}{x^2}
\pAb{}{x}\,\mathcal{D}(x)
\left[\pAb{}{x} \Delta n_{x}
+\xi(z) \Delta n_{x}\right],
\end{align}
where $x$ is the dimensionless frequency and $\xi(z)=\frac{h\nu_{\rm
    21}}{k\Te}\approx 40 \,\frac{1100}{[1+z]}$. The first term on the
right hand side describes photon diffusion and the second accounts for
the recoil effect.

\subsubsection{Electron scattering}

The diffusion coefficient in the case of electron scattering is
\citep[e.g. see][]{Zeldovich1969, Sazonov2000}
\beal
\label{app:D_e}
\mathcal{D}_{\rm e}(x)=\frac{\sigma_{\rm T} N_{\rm e} c} {H[1+z]}\, \left[\frac{k\Te }{\me c^2} \right] x^4,
\end{align}
where $\sigma_{\rm T}\approx \pot{6.65}{-25}\,\rm cm^2$ is the Thomson
cross section.  \citet{Chluba2009b} pointed out that electron
scattering has an effect only at the early stages of recombination
($z\gtrsim 1400$). However it is easy to include, and also has the
advantage of stabilizing the numerical treatment by damping small
scale fluctuations of the photon occupation number caused by numerical
errors, even in places where line scattering is already negligible.

\changeI{As can be seen from the form of the diffusion coefficient in Eq.~\eqref{app:D_e}, the efficiency of electron scattering to a large extent is {\it achromatic}. This is in stark contrast to the case of resonance scattering, which is most efficient only in a very narrow range around the line center (see next paragraph). Furthermore, the number of free electrons drops rapidly towards the end of recombination, such that the Fokker-Planck approximation is expected to break down \citep{Chluba2009b}. Nevertheless, the diffusion approximation remains sufficient for computations of the free electron fraction \citep[see][]{Yacine2010b}. }

\subsubsection{Resonance scattering}

For resonance scattering by a Lyman-$k$ line the diffusion coefficient is
\citep[e.g. see][and reference therein]{Basko1978a, Basko1978b,
  Rybicki2006, Chluba2009b}
\beal
\label{app:D_Ly-j}
\mathcal{D}_{k}(x)
\approx p^{k\rm p}_{\rm sc}\,\frac{\sigma^{k\rm p}_{\rm r} N_{\rm 1s} c} {H[1+z]}\, \left[\frac{k\Te }{m_{\rm H} c^2} \right]
\frac{\nu^2_{k1}}{\nu^2_{\rm 21}} x^2 \phi^{k\rm p}_{\rm V}(x),
\end{align}
where $\sigma^{k\rm p}_{\rm r}=\frac{3 \lambda^2_{k1}
}{8\pi}\,\frac{A_{k{\rm p1s}}}{\Delta\nu^{k\rm p}_{\rm D}}$ and
$\Delta\nu^{k\rm p}_{\rm D}$ denote the resonant-scattering cross
section and the Doppler width of the Lyman-$k$ resonance,
respectively.
For the Lyman-$\alpha$ line $\sigma^{\rm 2p}_{\rm r}\sim
\pot{1.91}{-13}\,\rm cm^2$ and $\Delta\nu^{\rm 2p}_{\rm D}\sim
\pot{2.35}{-5}\,\nu_{\rm 21}$ at $z\sim 1100$.  The Voigt profile
$\phi^{k\rm p}_{\rm V}(x)=\varphi^{k\rm p}_{\rm V}(x)\,\Delta\nu^{k\rm
  p}_{\rm D}$, is normalized as $\int_{-\infty}^{\infty} \phi^{k\rm
  p}_{\rm V}(x^{k\rm p}_{\rm D})\id x^{k\rm p}_{\rm D}=\int_0^{\infty}
\varphi^{k\rm p}_{\rm V}(\nu)\id \nu=1$. Where $x^{k\rm p}_{\rm
  D}=[\nu-\nu_{k \rm p}]/\Delta\nu^{k\rm p}_{\rm D}$ is the distance
to the line center in units of the Doppler width.

The scattering probability of the Lyman-$k$ resonance, $p^{k\rm
  p}_{\rm sc}$, is determined by a {\it weighted count} of all
possible ways {\it out} of the $k{\rm p}$-state, $R^{-}_{k\rm
  p}(\Tg)$, excluding the Lyman-series resonance being considered, and
then writing the branching ratio as\footnote{Stimulated emission for
  the Lyman-series has been neglected.}
\beal
\label{eq:p_sc_k}
p^{k\rm p}_{\rm sc}&=\frac{A_{k{\rm p1s}}}{A_{k{\rm p1s}}+R^{-}_{k\rm p}},
\end{align}
yielding the probability for re-injection into the Lyman-$k$
resonance.  The rates $R^{-}_{k{\rm p}}(\Tg)$ and the scattering
probabilities, $p^{k\rm p}_{\rm sc}(\Tg)$, can be pre-computed,
independent of the solutions obtained from the multi-level code. We
detail the procedure below.

Following \citet{RybickiDell94}, the diffusion coefficient is
$D\propto\phi^{k\rm p}_{\rm V}(\nu)$.  We neglect corrections due to non-resonant
contributions \citep[e.g. see][]{Lee2005} in calculating the
scattering cross section, which would lead to a different frequency
dependence far away from the resonance (e.g. Rayleigh scattering in the
distant red wing, \citealt{Jackson}).
However, because it turns out that resonance scattering {is} only
important in the vicinity of the Lyman-$\alpha$ resonance, this
approximation suffices.

\changeI{It is also worth mentioning that Eq.~\eqref{app:D_Ly-j} together with Eq.~\eqref{app:coll_n_nu_scatt}, in the limit of large optical depth\footnote{\changeI{During hydrogen recombination photons scatter efficiently off the Lyman-$\alpha$ resonance out to $\sim 10^4-10^5$ Doppler width \citep[see Fig.~3 in][]{Chluba2008b}. However, the redistribution of photons in the distant damping wings still remains rather slow \citep{Chluba2009b}.}}, provides a viable description for the redistribution of photon over frequency.
Unlike the case of {\it complete redistribution} (in which the reemission after each scattering event occurs over the whole Voigt profile), in the recombination epoch photons are only {\it partially redistributed} as a result of the Doppler motions of the hydrogen atoms, so-called type-II redistribution as defined in \citet{Hummer1962}.
}

\subsubsection{Equilibrium solution for the scattering term}
\label{sec:DNe_sc_eq}
Independent of the type of scattering, the equilibrium distribution
with respect to the scattering term Eq.~\eqref {app:coll_n_nu_scatt}
is given by
\beal
\label{eq:Dn_sc_eq}
\Delta n^{\rm sc,eq}_{x}= \Delta n_{x_0}(z) \, e^{-\xi(z)[x-x_0]}.
\end{align}
This is the expected Wien spectrum with the temperature defined by
the electrons. The normalization $\Delta n_{x_0}(z)$ is determined by
the emission and absorption process.

The optical depth to line scattering being extremely large inside the
Doppler cores of the Lyman-resonances ($\tau_{\rm S}\sim10^{6}-10^{8}$
during \ion{H}{i} recombination) causes the photon distribution within
the Doppler core to remain extremely close to equilibrium, $\Delta
n^{\rm sc,eq}_{x}$.

\subsection{Normal Lyman-$k$ emission and absorption terms}
\label{sec:Ly-n-emission}
In the normal '$1+1$' photon picture, the emission profile for each
Lyman-series resonance is given by a Voigt-profile, $\varphi^{k\rm
  p}_{\rm V}$, with Voigt-parameter $a^{k\rm p}$.
Given the rate, $R^+_{k\rm p}(\Tg, \Te)$, at which {\it
fresh}\footnote{Electrons that did not enter the p-state via the
  Lyman-$k$ resonance.} electrons reach the $k$p-state, and the
probability of photon injection into the Lyman-$k$ resonance, $p_{\rm
  em}^{k\rm p}\equiv p_{\rm sc}^{k\rm p}$, the Lyman-$k$ line-emission
and absorption term are \citep[e.g. see][]{Chluba2009};
\beal
\label{eq:Line_em_ab_k}
\left. \pAb{\Delta n_x}{z} \right|^{{\rm Ly-}k}_{\rm em/abs}
&=- p_{\rm d}^{k\rm p}\,\frac{\sigma^{k\rm p}_{\rm r} N_{\rm 1s} c} {H[1+z]}\,
\frac{\nu^2_{k1}}{\nu^2_{21}} \frac{\phi^{k\rm p}_{\rm V}}{x^2}
\left\{ \nu_{\rm 2p1s}\,\Delta n^{k\rm p}_{\rm em} - f^{k\rm p}_{x} \Delta n_x \right\}.
\end{align}
The factor $1/x^2$ accounts for the translation from photon number to
the occupation number because $\Delta N_\nu\propto \nu^2 \Delta
n_\nu$, for which the Voigt-profile is defined.
Also, $p_{\rm d}^{k\rm p}=1-p_{\rm em}^{k\rm p}$ is the death or the
real absorption probability in the $k^{\rm th}$ Lyman-series
resonance, and $\Delta n^{k\rm p}_{\rm em}$ and $f^{k\rm p}_{x}$ are
given by,
\bsub
\label{eq:Dn_em_k}
\beal
\Delta n^{k\rm p}_{\rm em}
&=\frac{g_{\rm 1s}}{g_{k\rm p}}\frac{R^+_{k\rm p}}{R^-_{k\rm p}\, N_{\rm 1s}}-e^{-{h\nu}_{k1}/k\Tg}
\\
f^{k\rm p}_{x}&=\exp\left(h[\nu-\nu_{k1}]/k\Tg\right),
\end{align}
\esub
where $g_{\rm 1s}/g_{k\rm p}$ {is} the ratio of the statistical
weights of the initial and final states.
The function $\Delta n^{k\rm p}_{\rm em}(\Tg,\Te)$ can in principle be
pre-computed using the solution for the populations of the levels from
the initial run of the effective multi-level recombination code.
However, the simplest way to define the ratio $R^+_{k\rm p}/R^-_{k\rm
  p}$ is to use the quasi-stationary approximation for the
$n$p-population (see details below).
We note that in full thermodynamic equilibrium $\Delta n^{k\rm p}_{\rm em}=0$, so that no distortion is created ($\Delta n_x=0$).

Physically, Eq.~\eqref{eq:Line_em_ab_k} includes two important
aspects, which are not considered in the standard recombination
calculation. Firstly, it allows for a distinction between scattering
events on one side, and real emission and absorption events on the
other. Secondly, it ensures conservation of blackbody spectrum in full
thermodynamic equilibrium, even in the very distant wings of the
lines. Refer \citet{Chluba2009} for a detailed explanation of the latter
point, and on how this leads to one of the largest corrections in the
case of Lyman-$\alpha$ transport.

\subsubsection{Computing $\Delta n_{\rm em}^{k\rm p}$}
\label{sec:Dnem_kp}
To solve the evolution of the photon field, one has to know
at which rate photons are produced by the Lyman-resonance.
This rate depends on $\Delta n^{k\rm p}_{\rm em}$ as defined in Eq.~\eqref{eq:Dn_em_k}.

The rate equation for the evolution of the population in the $k$p-level has the form
\citep[see Appendix B][]{Chluba2009},

\bsub
\label{eq:dNnpdt}
\beal
\Abl{X_{k\rm p}}{t}
&=\left.\Abl{X_{k \rm p}}{t}\right|_{{\rm Ly}-k}+R^{+}_{k\rm p}-R^{-}_{k\rm p}X_{k\rm p}
\\
\left.\Abl{X_{k \rm p}}{t}\right|_{{\rm Ly}-k}
&=\frac{g_{\rm 1s}}{g_{k \rm p}}
A_{k\rm p1s}\,X_{\rm 1s} \, \mathcal{I}^{k\rm p}_1
- A_{k\rm p1s}\,X_{k\rm p}\mathcal{I}^{k\rm p}_2
\\
\label{eq:dNnpdt_c}
\mathcal{I}^{k\rm p}_1&= \int \varphi^{k\rm p}_{\rm V}(\nu) \, e^{h[\nu-\nu_{k1}]/k\Tg}\,n_\nu \id\nu
\\
\mathcal{I}^{k\rm p}_2&= \int \varphi^{k\rm p}_{\rm V}(\nu) [1+n_\nu]\id\nu\approx 1+\nbb(\nu_{k1})\approx 1.
\end{align}
\esub
In this picture the emission, absorption and resonance scattering
terms are all treated simultaneously. In addition, the asymmetry
between the emission and absorption profile in the Lyman-$k$
resonance, as required by detailed balance, has been incorporated.

Under quasi-stationarity, and using the definition of the death
probability, $p^{k\rm p}_{\rm d}$, Eq.~\eqref{eq:dNnpdt} yields
\beal
\label{eq:Rpkp_Rmkp}
\frac{g_{\rm 1s}}{g_{k\rm p}}\,\frac{R^{+}_{k\rm p}}{R^{-}_{k\rm p}\,X_{\rm 1s}}
=\frac{1}{p^{k\rm p}_{\rm d}}\left[\frac{g_{k\rm p}}{g_{\rm 1s}}\,\frac{X_{k\rm p}}{X_{\rm 1s}}-\mathcal{I}^{k\rm p}_1\right] 
+\mathcal{I}^{k\rm p}_1,
\end{align}
{such that with} Eq.~\eqref{eq:Dn_em_k}
\bsub
\label{eq:Dnem_kp1s}
\beal
\label{eq:Dnem_kp1s_a}
\Delta n^{k\rm p}_{\rm em}
&= \frac{1}{p^{k\rm p}_{\rm d}}\left[\frac{g_{k\rm p}}{g_{\rm 1s}}\,\frac{X_{k\rm p}}{X_{\rm 1s}}-\mathcal{I}^{k\rm p}_1\right] 
+\mathcal{I}^{k\rm p}_1 - e^{-{h\nu}_{k\rm p1s}/k\Tg}
\\
\label{eq:Dnem_kp1s_b}
&\approx \Delta n^{k }_{\rm L} \left[1+\frac{p^{k\rm p}_{\rm em}}{p^{k\rm p}_{\rm d}} P^{k}_{\rm S} \right].
\end{align}
\esub
{In the second step we used} the normal Sobolev approximation, for which
$\mathcal{I}^{k\rm p}_1\approx n^{k }_{\rm L}-P^{k}_{\rm S}\left[n^{k
  }_{\rm L}-n^{\rm pl}_{k\rm p1s}\right]$ \citep[for the case of
  Lyman-$\alpha$ compare also with Eq. (41) in][]{Chluba2008b}.

From Eq.~\eqref{eq:Dnem_kp1s_b} we have $\Delta n^{k\rm
  p}_{\rm em} \approx\Delta n^{k }_{\rm L}$, since for all
Lyman-series resonances the second term in brackets is very
small. Nevertheless, for the total normalization of the line intensity
close to the line center, this small correction is important
\citep{Chluba2008b}, in particular for the Lyman-$\alpha$ resonance.

Also we would like to mention that for the Voigt parameter of the
Lyman-$k$ profiles, $a^{k\rm p}=A^{k\rm p}_{\rm
  tot}/[4\pi\Delta\nu^{k\rm p}_{\rm D}]$, the {\it total} width of the
line is used, where transitions induced by the CMB blackbody (e.g. to
higher levels) are included.  Numerically, it is {possible} to compute
the total width for the Lyman-$k$ resonance with $A^{k\rm p}_{\rm
  tot}\equiv A_{k\rm p 1s}/p^{{k\rm p}}_{\rm em}$.

\subsection{The 2s-1s two-photon channel}
The 2s-1s two-photon channel provides the pathway for about 60\% of
all electrons in hydrogen to settle into the ground state
\citep{Chluba2006b}. It therefore provides the most important
{channel} in the cosmological recombination process.
Here we treat the case of 2s-1s separately to illustrate the
important approximations in the two-photon picture. The derivation
outlined in this section is then used to obtain the corresponding
terms for the two-photon processes from excited states with $n>2$ (see
\S~\ref{sec:two-gamma-term}).

The net change of the number density of electrons in the 2s level via the
2s-1s two-photon channel is given by
\beal
\label{eq:dN2sdt_2g}
\left.\Abl{X_{\rm 2s}}{t}\right|^{2\gamma}_{\rm 1s}&=A^{2\gamma}_{\rm 2s1s}X_{\rm 1s}\int
\varphi^{2\gamma}_{\rm 2s}n(\nu)\,n(\nu_{21}-\nu)\id\nu
\nonumber
\\
&\quad
-A^{2\gamma}_{\rm 2s1s}X_{\rm 2s}\int \varphi^{2\gamma}_{\rm 2s} [1+n(\nu)] [1+n(\nu_{21}-\nu)]\id\nu,
\end{align}
where $A^{2\gamma}_{\rm 2s1s}=8.2206\,\text{s}^{-1}$
\citep{Labzowsky2005} is the vacuum 2s-1s two-photon decay rate, and
$\varphi^{2\gamma}_{\rm 2s}$ denotes the 2s-1s two-photon decay
profile normalized as $\int \varphi^{2\gamma}_{\rm 2s}\id\nu=1$.
Including all possible ways in and out of the 2s-level the net change of the
number density of electrons in the 2s-state can be written as
\beal
\label{eq:dN2sdt}
\Abl{X_{\rm 2s}}{t}=\left.\Abl{X_{\rm 2s}}{t}\right|^{2\gamma}_{\rm 1s}+R^{+}_{\rm 2s}-R^{-}_{\rm 2s}X_{\rm 2s}.
\end{align}
Here $R^{+}_{\rm 2s}$ and $R^{-}_{\rm 2s}$ include the effect of all transitions to bound states with $n>2$ and the continuum. 

In order to simplify the notation we now introduce
\bsub
\label{eq:2saverage}
\beal
\label{eq:2saverage_a}
\twosavg{f(\nu)}{2\gamma}{i}&=\int_0^{\nu_{i\rm 1s}} \varphi^{2\gamma}_{i} f(\nu) \id\nu,
\\
\mathcal{G}^{i}_1&=\twosavg{n\,n'}{2\gamma}{i}
\\
\label{eq:2saverage_c}
\mathcal{G}^{i}_2&=\twosavg{[1+n][1+n']}{2\gamma}{i}
\end{align}
\esub
where $f(\nu)$ is some arbitrary function of frequency and $n=n(\nu)$ and
$n'=n(\nu')$ with $\nu'=\nu_{i1\rm s}-\nu$.

Then, under {\it quasi-stationarity} the solution for the
population of the 2s-state is given by
\beal
\label{eq:N2sQS}
X_{\rm 2s}^{\rm QS}=
\frac{R^{+}_{\rm 2s}+A^{2\gamma}_{\rm 2s1s}X_{\rm 1s}\,\mathcal{G}^{\rm 2s}_1}
{R^{-}_{\rm 2s}+A^{2\gamma}_{\rm 2s1s}\,\mathcal{G}^{\rm 2s}_2}.
\end{align}
In the multi-level approach the effect of stimulated two-photon
emission is neglected leading to $\mathcal{G}^{\rm 2s}_2\approx
1$. Also any CMB spectral distortion that is introduced by the
recombination process {(e.g. because of
Lyman-$\alpha$ emission) is omitted}, implying $\mathcal{G}^{\rm 2s}_1\approx
\twosav{n^{\rm pl}\,{n^{\rm pl}}'}\approx \exp(-h\nu_{\rm
  21}/k\Tg)$. In this approximation, the result from
Eq.~\eqref{eq:N2sQS} becomes identical to the one obtained using
Eq.~\eqref{eq:pert_X1s_stand_b} and Eq.~\eqref{eq:dN2sdt}, in the
standard multi-level approach.

However, in the recombination problem corrections to both
$\mathcal{G}^{\rm 2s}_1$ and $\mathcal{G}^{\rm 2s}_2$ are important.
For the stimulated two-photon emission {only} the occupation number given by the undistorted CMB blackbody has to be considered and thus, 
\beal
\label{eq:I2s_2_approx}
\mathcal{G}^{\rm 2s}_2&\approx \twosav{[1+\nbb][1+\nbb']}\equiv \mathcal{G}^{\rm 2s, pl}_2, 
\end{align}
which can be precomputed as a function of temperature. Typically, $\mathcal{G}^{\rm 2s, pl}_2$ exceeds unity by a few percent \citep{Chluba2006}.

For $\mathcal{G}^{\rm 2s}_1$ one can make use of the fact that the distortions at either $\nu$ or $\nu'$ are very small, so that
\beal
\label{eq:npnpp}
n\,n' \approx n^{\rm pl}\,{n^{\rm pl}}' + {n^{\rm pl}}'\,\Delta n+ {n^{\rm pl}}\,\Delta n'.
\end{align}
Hence Eq.~\eqref{eq:dN2sdt_2g} can be re-written as,
\bsub
\label{eq:dN2sdt_2g_tilde}
\beal
\label{eq:dN2sdt_2g_tilde_a}
\left.\Abl{X_{\rm 2s}}{t}\right|^{2\gamma}_{\rm 1s}
&=A^{2\gamma, \ast}_{\rm 2s1s}\left[ X_{\rm 1s} e^{-h\nu_{\rm 21}/k\Tg} - X_{\rm 2s} \right]
+ A^{2\gamma}_{\rm 2s1s}\,X_{\rm 1s}\,\Delta \mathcal{G}^{\rm 2s}_1
\\[1mm]
\label{eq:dN2sdt_2g_tilde_b}
\Delta \mathcal{G}^{\rm 2s}_1
&=\!\!
\int
\!\varphi^{2\gamma}_{\rm 2s}\left[{n^{\rm pl}}' \Delta n+ {n^{\rm pl}}\,\Delta n'\right]\! \id\nu
\equiv\!
2\!\int^{\nu_{21}}_{\frac{\nu_{21}}{2}}
\!\!\varphi^{2\gamma}_{\rm 2s}\,{n^{\rm pl}}' \Delta n \id\nu,
\end{align}
\esub
where we defined the stimulated 2s-1s two-photon decay rate within the
CMB ambient radiation field as $A_{\rm 2s1s}^{2\gamma,
  \ast}=A^{2\gamma}_{\rm 2s1s}\,\mathcal{G}^{\rm 2s, pl}_2$
\citep[cf.][]{Chluba2006}.
Also Eq.~\eqref{eq:dN2sdt_2g_tilde_b} reflects the symmetry of the
two-photon profile around $\nu=\nu_{21}/2$.

Note that for $\mathcal{G}^{\rm 2s, pl}_2$ only the CMB blackbody
spectrum is important and therefore can, in principle, be precomputed as
a function of photon temperature, $\Tg$.
\changeI{This also emphasizes the difference in the origin of the two terms of Eq.~\eqref{eq:dN2sdt_2g_tilde_a}, $\mathcal{G}^{\rm 2s, pl}_2$ being the {\it thermal} contribution, while $\Delta \mathcal{G}^{\rm 2s}_1$ arises solely because of {\it non-thermal} photons created in the recombination process.}

By comparing Eq.~\eqref{eq:dN2sdt_2g_tilde} with
Eq.~\eqref{eq:pert_X1s_stand_b} one can write down the correction to
the 2s-1s net two-photon rate
\beal
\label{eq:2s_1s_correct}
\Delta R^{\rm corr}_{\rm 2s \leftrightarrow 1s}&=
A^{2\gamma}_{\rm 2s1s}\Delta \mathcal{G}^{\rm 2s, pl}_2 
\left[ X_{\rm 1s} e^{-h\nu_{\rm 21}/k\Tg} - X_{\rm 2s} \right]
+ A^{2\gamma}_{\rm 2s1s}\,X_{\rm 1s}\,\Delta \mathcal{G}^{\rm 2s}_1.
\end{align}
%
Here we introduced $\Delta \mathcal{G}^{\rm 2s, pl}_2=\mathcal{G}^{\rm
  2s, pl}_2-1$, which during recombination is of order $\sim 1\%$.
In Equation~\eqref{eq:dN2sdt_2g_tilde} the integral $\Delta
\mathcal{G}^{\rm 2s}_1$ depends on the spectral distortion introduced
by the recombination process in the Wien's tail of the CMB blackbody.
Including only the Lyman-$\alpha$ distortion provides a manner in
which to take its feedback effect into account
\citep[cf.][]{Kholu2006}.

\subsubsection{The 2s-1s two-photon emission and absorption term}
In contrast to the Lyman-series channels, the terms for the photon
radiative transfer equation in the case of the 2s-1s channel can be {directly}
obtained from the net rate between the 2s and 1s state as in
Eq.~\eqref{eq:dN2sdt_2g}, resulting in

\beal
\label{eq:dnnu_2sdt_2g}
\left.\frac{1}{c}\,\frac{\partial N_{\nu}}{\partial t}\right|^{\rm 2s1s}_{2\gamma}
&=A^{2\gamma}_{\rm 2s1s}N_{\rm 2s} \tilde{\varphi}^{2\gamma}_{\rm 2s} [1+n(\nu)] [1+n(\nu_{21}-\nu)]
\nonumber
\\
&\qquad\qquad
-A^{2\gamma}_{\rm 2s1s}N_{\rm 1s} \tilde{\varphi}^{2\gamma}_{\rm 2s}n(\nu)\,n(\nu_{21}-\nu).
\end{align}
Here we defined $\tilde{\varphi}^{2\gamma}_{\rm
  2s}=\frac{2\,\varphi^{2\gamma}_{\rm 2s}}{4\pi}$, where the factor of
two results from two photons being added to the photon field, and the
$4\pi$ converts the units to per steradian.

The reason for this simple connection to the net rate equation is
related to the fact that every transition from the 1s state to the 2s
level is expected to lead to a {\it complete redistribution} over the
2s-1s two-photon profile. The main reason behind this assumption of
redistribution is that the probability of coherent 1s-2s {\it
  scattering} event is tiny because the 2s-1s decay rate is extremely
small compared to the time it takes to excite a 2s-electron to higher
levels or the continuum.

However, some additional simplifications are possible. First, we can
again replace the factors, $[1+n][1+n']$, accounting for stimulated
two-photon emission with those from the undistorted CMB
blackbody. Furthermore, from Eq.~\eqref{eq:npnpp}, 
\beal
\label{eq:npnpp_2}
n\,n' \approx e^{-h\nu_{\rm 21}/k\Tg}\,[1+ n^{\rm pl}][1+{n^{\rm pl}}']\left[1 + \frac{\Delta n}{n^{\rm pl}}+ \frac{\Delta n'}{{n^{\rm pl}}'}\right].
\end{align}
Also, since the spectral distortions at very low frequencies are never
important, one of the two terms in Eq.~\eqref{eq:npnpp_2} (say the one
related to $\Delta n'$) can always be omitted.
Therefore we can rewrite Eq.~\eqref{eq:dnnu_2sdt_2g} as
\beal
\label{eq:dNdt_2s_emabs_tilde}
\left.\frac{1}{c}\,\frac{\partial N_{\nu}}{\partial t}\right|^{\rm 2s1s}_{2\gamma}
&=A^{2\gamma}_{\rm 2s1s} N_{\rm 1s}\,\tilde{\varphi}^{2\gamma, \ast}_{\rm 2s}
\left[\Delta n_{\rm em}^{\rm 2s} - f^{\rm 2s}_\nu \Delta n_\nu \right]
\end{align}
where $\tilde{\varphi}^{2\gamma, \ast}_{\rm 2s}\equiv \tilde{\varphi}^{2\gamma}_{\rm 2s}[1+\nbb][1+\nbb']$ and
\bsub
\label{eq:Dnem}
\beal
\label{eq:Dnem_a}
\Delta n_{\rm em}^{\rm 2s}&=\frac{X_{\rm 2s}}{X_{\rm 1s}}- e^{-h\nu_{\rm 21}/k\Tg}
\\
\label{eq:Dnem_b}
f^{\rm 2s}_\nu &= \frac{e^{-h\nu_{\rm 21}/k\Tg}}{n^{\rm pl}(\nu)}\approx \exp\left(h[\nu-\nu_{\rm 21}]/k\Tg\right).
\end{align}
\esub
If the term $\frac{\Delta n'}{{n^{\rm pl}}'}$ is {non-negligible}, as
might be the case at very low redshifts ($z\lesssim 400$), where the
Lyman-$\alpha$ photons emitted at $z\sim 1400$ {redshifts} into the 2s-1s
absorption channel, one in addition has to subtract the term $f^{\rm
  2s}_{\nu'} \Delta n_{\nu'}$ {within} the brackets of
Eq.~\eqref{eq:dNdt_2s_emabs_tilde}.
In terms of $x=\nu/\nu_{\rm 21}$, $z$ and $\Delta n_x$ the photon occupation
number now evolves as,
\beal
\label{eq:dDn_xdt_2s_emabs_tilde}
\left.\frac{\partial \Delta n_x}{\partial z}\right|^{\rm 2s1s}_{2\gamma}
&=
- \frac{\sigma^{2\gamma}_{\rm 2s1s} N_{\rm 1s} c}{H[1+z]}\,\frac{\phi^{2\gamma, \ast}_{\rm 2s}}{x^2}
\left[\nu_{\rm 21}\,\Delta n_{\rm em}^{\rm 2s} - f^{\rm 2s}_x \Delta n_x \right].
\end{align}
Here the 2s-1s cross section is given by $\sigma^{2\gamma}_{\rm
  2s1s}=\frac{\lambda^2_{21} A^{2\gamma}_{\rm 2s1s}}{8\pi \nu_{21}}$,
and $\phi^{2\gamma, \ast}_{\rm
  2s}=4\,\pi\,\nu_{21}\tilde{\varphi}^{2\gamma, \ast}_{\rm 2s}$.

Eq.~\eqref{eq:dDn_xdt_2s_emabs_tilde} bears a striking resemblance to
the emission and absorption in the Lyman-series channels as in
Eq.~\eqref{eq:Line_em_ab_k} because one of the two photons that are
involved in the 2s-1s two-photon process is drawn from the undistorted
CMB blackbody spectrum, so that the evolution equation essentially
\changeI{becomes} a {\it one-photon} equation.
The difference is the absence of a death probability since {practically} every
electron that is excited to the 2s state will take a detour to higher
levels or the continuum {as} $p^{\rm 2s}_{\rm d}\approx 1$.

\subsection{Two-photon emission and absorption terms from excited levels with $n>2$}
\label{sec:two-gamma-term}
One of the most interesting modifications to the solution for the
photon field {is related to the deviations of the profiles for the different two-photon emission and absorption channels from the Lorentzian shape \citep{Chluba2008a}.}
For the recombination problem only those one-photon sequences
{involving} a Lyman-series resonance (e.g. ${\rm 4d}\leftrightarrow{\rm
  2p}\leftrightarrow{\rm 1s}$) are important\footnote{All the other
  two-photon emission and absorption channel (e.g. ${\rm
    4d}\leftrightarrow{\rm 2p}\leftrightarrow{\rm 2s}$) can be treated
  within a blackbody ambient radiation field, so that their net rate can
  be directly computed. Without deviations from the blackbody shape
  these will be extremely close to the normal '$1+1$' photon
  rates. Also they can only affect the net recombination rate as a
  'correction to correction', because they only act on the electron
  'feeding rates' into the main channels towards the ground state. A
  similar argument holds for Raman scattering events that do not
  directly connect to the ground state.}.
In this section we shall replace the standard '$1+1$'-photon terms
for these channels with the full two-photon description that takes
into account the coherent nature of the process\footnote{Conditions
  persistent in the Universe at the recombination epoch makes
  collisions negligible, maintaining the coherence of the two-photon
  decay \citep[e.g. see][]{Chluba2008a, Hirata2008}}.

We generalize the approach detailed in \citet{Chluba2009} for emission
of photons close to the Lyman-$\alpha$ line to include corrections
around the Lyman-$\beta$ and higher resonances.

\subsubsection{Net rates for two-photon transitions from excited s- and d-states}
The net change of the number density of electrons in the level
$j~\in~\{n{\rm s}, n{\rm d}\}$ via the $j$-1s two-photon channel is
given by
\beal
\label{eq:dNjdt_2g}
\left.\Abl{X_{j}}{t}\right|^{2\gamma}_{\rm 1s}&=\frac{g_j}{g_{\rm 1s}}A^{2\gamma}_{j\rm 1s}X_{\rm 1s}\int
\varphi^{2\gamma}_{j} (\nu)\, n(\nu)\,n(\nu_{j1}-\nu)\id\nu
\nonumber
\\
&\quad
-A^{2\gamma}_{j\rm 1s}X_{j}\int \varphi^{2\gamma}_{j} (\nu) \,[1+n(\nu)] [1+n(\nu_{j1}-\nu)]\id\nu.
\end{align}
{Here} $\varphi^{2\gamma}_{j}$ denotes the profile for the $j$-1s
two-photon decay, which can be computed as explained in
Appendix~\ref{app:comp_two_photon_profs}, and is
normalized\footnote{Small correction to the normalization due to the
  two-photon description are neglected.} as $\int
\varphi^{2\gamma}_{j} \id\nu=1$.
The (vacuum) two-photon decay rate is given by
\beal
\label{eq:A2gamma_j1s}
A^{2\gamma}_{j\rm 1s}&=\sum_{k=2}^{n_j-1} A_{j\,k{\rm p}} \, p^{k{\rm p}}_{\rm em}.
\end{align}
The ratio of the statistical weights is $g_j/g_{\rm 1s}=1$ for the
$n$s-states, and $g_j/g_{\rm 1s}=5$ for $n$d-states.
Equation~\eqref{eq:A2gamma_j1s} simply reflects the one-photon decay
rates and branching ratios of all the '$1+1$' photon routes
$j\rightarrow n{\rm p}\rightarrow {\rm 1s}$ via intermediate p-states
with $n<n_j$.
Stimulated emission induced by the CMB photons is not included in
the definition of $A^{2\gamma}_{j\rm 1s}$, since it is taken into
account differentially by the integrals in Eq.~\eqref{eq:dNjdt_2g}.

With notations defined in Eq.~\eqref{eq:2saverage}, and following the
procedure to derive Eq.~\eqref{eq:dN2sdt_2g_tilde}, we can re-write
Eq.~\eqref{eq:dNjdt_2g} as
\bsub
\label{eq:dNjdt_2g_tilde}
\beal
\label{eq:dNjdt_2g_tilde_a}
\left.\Abl{X_{j}}{t}\right|^{2\gamma}_{\rm 1s}
&=A^{2\gamma, \ast}_{j\rm 1s}\left[ \frac{g_j}{g_{\rm 1s}}\,X_{\rm 1s} e^{-h\nu_{j\rm 1s}/k\Tg} - X_{j} \right]
+ \frac{g_j}{g_{\rm 1s}}\, A^{2\gamma}_{j \rm 1s}\,X_{\rm 1s}\,\Delta \mathcal{G}^{j}_1
\\[1mm]
\label{eq:dNjdt_2g_tilde_b}
\Delta \mathcal{G}^{j}_1
&=
2\int^{\nu_{j\rm1s}}_{\nu_{j\rm1s}/2}
\varphi^{2\gamma}_{j}\,{n^{\rm pl}}' \Delta n \id\nu,
\end{align}
\esub
where we defined the stimulated $j$-1s two-photon decay rate within
the CMB ambient radiation field as $A_{j \rm 1s}^{2\gamma,
  \ast}=A^{2\gamma}_{j\rm 1s}\,\mathcal{G}^{j\rm, pl}_2$.

The $\mathcal{G}^{j\rm, pl}_2$ \changeI{depends crucially only} on the CMB
blackbody spectrum and thus can be \changeI{precomputed} as a function of \changeI{photon}
temperature, $\Tg$.
\changeI{On the other hand, like for the 2s-1s two-photon process (see Eq.~\eqref{eq:dN2sdt_2g_tilde}), $\Delta \mathcal{G}^{j}_1$ arises due to {\it non-thermal} photons, and hence depends directly on the solution for the photon field.}

In the normal '$1+1$' photon picture, the two-photon profiles can be
considered as a sum of $\delta$-functions and therefore
\beal
\label{eq:A_2g_stim}
A_{j \rm 1s}^{2\gamma(1+1), \ast}= \sum_{k=2}^{n_j-1} A^\ast_{j\,k{\rm p}} \, p^{k{\rm p}}_{\rm em}.
\end{align}
Here $A^\ast_{j\,k{\rm p}}=A_{j\,k{\rm p}}[1+\nbb(\nu_{jk})]$, and the
stimulated effect close to the Lyman-series resonances has been
neglected, i.e. $1+\nbb(\nu_{k\rm p1s})\approx 1$.

\subsubsection{Two-photon emission and absorption for excited s- and d-states}
The two-photon emission and absorption terms are obtained following
the steps in the derivation of Eq.~\eqref{eq:dDn_xdt_2s_emabs_tilde}. 
For the $j$-1s two-photon channel one therefore obtains
\beal
\label{eq:dDn_xdt_j_emabs_tilde}
\left.\frac{\partial \Delta n_x}{\partial z}\right|^{j\rm 1s}_{2\gamma}
&=
-\frac{\sigma^{2\gamma}_{j\rm 1s} N_{\rm 1s} c}{H[1+z]}\,\frac{\nu^2_{j1}}{\nu^2_{21}}\,\frac{\phi^{2\gamma, \ast}_{j}}{x^2}
\left[\nu_{\rm 21}\,\Delta n_{\rm em}^{j} - f^{j}_x \Delta n_x \right].
\end{align}
The $j$-1s two-photon cross section is given by $\sigma^{2\gamma}_{j
  \rm 1s}=\frac{g_j}{g_{\rm 1s}}\,\frac{\lambda^2_{j1}
  A^{2\gamma}_{j\rm 1s}}{8\pi \nu_{j1}}$, and $\phi^{2\gamma,
  \ast}_{j}=4\,\pi\,\nu_{j1}\tilde{\varphi}^{2\gamma}_{j}(\nu)\,[1+\nbb(\nu)][1+\nbb(\nu')]$, where, because of energy conservation, $\nu'=\nu_{j\rm 1s}-\nu$.
Also,
\bsub
\label{eq:Dnem_j}
\beal
\label{eq:Dnem_j_a}
\Delta n_{\rm em}^{j}&=\frac{g_{\rm 1s}}{g_j}\frac{X_{j}}{X_{\rm 1s}}- e^{-h\nu_{j1}/k\Tg}
\\
\label{eq:Dnem_j_b}
f^{j}_\nu &= \frac{e^{-h\nu_{j1}/k\Tg}}{n^{\rm pl}(\nu)}\equiv \frac{e^{h[\nu-\nu_{j1}]/k\Tg}}{1+n^{\rm pl}(\nu)}
\approx \exp\left(h[\nu-\nu_{j1}]/k\Tg\right).
\end{align}
\esub
Again we emphasize the resemblance of the equation above to that of the
one-photon equation for the Lyman-series emission and absorption
channels as in Eq.~\eqref{eq:Line_em_ab_k}.

\subsubsection{Correcting the Lyman-series emission and absorption terms in the radiative transfer equation}
\label{sec:double_1}
Two-photon decays from a given initial state $j~\in~\{n{\rm s}, n{\rm
  d}\}$ involve Lyman-series resonances with $k<n$. For example, a
4d-1s two-photon emission event includes the effect of the
Lyman-$\alpha$ and $\beta$ resonance.
In the Lyman-series emission and absorption terms as in
Eq.~\eqref{eq:Line_em_ab_k}, these are already accounted for as
'$1+1$' photon terms, when the profile is given by the normal Voigt
function.

To avoid the {\it double counting} of these transitions in the
radiative transfer equation, two modifications are necessary: (i) all
death probabilities, $p^{k\rm p}_{\rm d}$, have to be reduced to
account only for those channels that are not included in the two-photon
description, and (ii) the Lyman-series emission rates have to be
reduced for the same reason.
This approach was also explained in \citet{Chluba2009} for the 3s-1s
and 3d-1s two-photon process.
Including {\it only} the $j$-1s two-photon process (say for 3d-1s),
the modified death probability and $\Delta \tilde{n}_{\rm em}^{k\rm
  p}$ of the Lyman-$k$ resonance becomes,
\bsub
\label{eq:mod_pd_Dnem_j}
\beal
\label{eq:mod_pd_Dnem_j_a}
\tilde{p}^{k\rm p}_{\rm d} &= p^{k\rm p}_{\rm d} - p^{j, k\rm p}_{\rm d}
\\
\label{eq:mod_pd_Dnem_j_b}
\Delta \tilde{n}_{\rm em}^{k\rm p}
&= \frac{1}{3\, X_{\rm 1s}}\, \frac{R^{+}_{k\rm p}-R^{j, +}_{k\rm p}}{R^{-}_{k\rm p}-R^{j, -}_{k\rm p}} - e^{-h\nu_{k\rm p}/k\Tg}
%
\end{align}
where 
%
%
the partial death probability, $p^{j, k\rm p}_{\rm d}$, is given by
\beal
\label{eq:mod_pd_Dnem_j_c}
p^{j, k\rm p}_{\rm d}&= \frac{R^{j, -}_{k\rm p}}{A_{k{\rm p1s}}+R^{-}_{k\rm p}} 
\equiv  p^{k\rm p}_{\rm em}\,\frac{R^{j, -}_{k\rm p}}{A_{k{\rm p1s}}}
\equiv p^{k\rm p}_{\rm d}\,\frac{R^{j, -}_{k\rm p}}{R^{-}_{k\rm p}}. 
\end{align}
{The} partial rates in and out of the $k$p-state are
\beal
\label{eq:mod_pd_Dnem_j_d}
R^{j, +}_{k\rm p}
&=A_{j \, k{\rm p}}[1+\nbb(\nu_{j\, k{\rm p}})] \, X_j
\\
\label{eq:mod_pd_Dnem_j_e}
R^{j, -}_{k\rm p}&=\frac{g_j}{g_{k\rm p}}\,A_{j \, k{\rm p}}\,\nbb(\nu_{j\, k{\rm p}})
\end{align}
\esub
such that $\frac{g_{\rm 1s}}{g_{k\rm p}}\frac{R^{j, +}_{k\rm p}}{R^{j,
    -}_{k\rm p}\,X_{\rm 1s}}\equiv n^j_{\rm L}\,\,e^{h\nu_{j\,k\rm
    p}/k\Tg}$ with $n^j_{\rm L}=\frac{g_{\rm 1s}}{g_j}
\frac{X_j}{X_{\rm 1s}}$.

When more than one two-photon channel is included, then for every
Lyman-resonance the following needs to be computed;
\bsub
\label{eq:mod_pd_Dnem_j_all}
\beal
\tilde{p}^{k\rm p}_{\rm d} &= p^{k\rm p}_{\rm d} - \sum_j \,p^{j, k\rm p}_{\rm d}
\\
\Delta \tilde{n}_{\rm em}^{k\rm p}
&= \frac{1}{3\, X_{\rm 1s}}\, \frac{R^{+}_{k\rm p}-\sum_j \,R^{j, +}_{k\rm p}}{R^{-}_{k\rm p}-\sum_j \,R^{j, -}_{k\rm p}} - e^{-h\nu_{k\rm p}/k\Tg},
%
\end{align}
\esub
where the sums run over all involved initial levels $j$.

\subsubsection{Correcting the net rates in the multi-level atom}
\label{sec:corr_2g}
\label{sec:double_2}
Equation~\eqref{eq:dNjdt_2g} relates the population of level $j$ with
the ground state.
The corresponding net two-photon transition rate includes the effect
of all '$1+1$' photon processes, $j\leftrightarrow n{\rm
  p}\leftrightarrow {\rm 1s}$, via Lyman-series resonances with
$n<n_j$.
Double-counting can again be avoided by subtracting the corresponding
'$1+1$' photon terms from the full $j$-1s two-photon rate.
The remaining corrections can then be added to the effective
multi-level code as additional rates which directly connects level $j$
to the ground state\footnote{ \citet{Chluba2009} proposed a varied
  treatment in which the '$1+1$' photon terms were first taken out of
  the standard network of rate equations and then the full two-photon
  rate between level $j$ and 1s added, which at the end, is completely
  equivalent.
}.

In the standard multi-level description of all $j\leftrightarrow
n{\rm p}\leftrightarrow {\rm 1s}$ sequences ($n_j>n$), the
contributions to the two-photon net rate as in Eq.~\eqref{eq:dNjdt_2g},
take the form \citep[see also][]{Chluba2009}
\bsub
\label{eq:dNjdt_1p1}
\beal
\label{eq:dNjdt_1p1_a}
\left.\Abl{X_{j}}{t}\right|^{2\gamma (1+1)}_{{\rm 1s},k{\rm p}}
&=
\frac{g_{k\rm p}}{g_{\rm 1s}}X_{\rm 1s}\,A_{k{\rm p1s}} \, p^{j, k\rm p}_{\rm d}\,\bar{n}_{k{\rm p1s}}
-X_{j}\, A^\ast_{j\,k{\rm p}} \, p^{k{\rm p}}_{\rm em}
\\
\label{eq:dNjdt_1p1_b}
&\equiv
\frac{g_j}{g_{\rm 1s}}X_{\rm 1s}\,A_{j\,k{\rm p}} \, p^{k{\rm p}}_{\rm em}\,\nbb(\nu_{jk})\,\bar{n}_{k{\rm p1s}}
-X_{j}\, A^\ast_{j\,k{\rm p}} \, p^{k{\rm p}}_{\rm em}
\\
\label{eq:dNjdt_1p1_c}
\left.\Abl{X_{j}}{t}\right|^{2\gamma (1+1)}_{\rm 1s}
&=
\sum_{k=2}^{n_j-1} \left.\Abl{X_{j}}{t}\right|^{2\gamma (1+1)}_{{\rm 1s},k{\rm p}}.
\end{align}
\esub
Equation~\eqref{eq:dNjdt_1p1_a} is interpreted as electrons exiting level
$j$ via the route $j\rightarrow n{\rm p}\rightarrow {\rm 1s}$ at a
rate $ A^\ast_{j\,k{\rm p}}$ times the probability, $p^{k{\rm p}}_{\rm
  em}$ {(second term). Similarly}, electrons reach state $j$ from
the ground state via the route ${\rm 1s}\rightarrow n{\rm
  p}\rightarrow j$, with the Lyman-$k$ excitation rate, $\frac{g_{k\rm
    p}}{g_{\rm 1s}}\,A_{k{\rm p1s}}\,\bar{n}_{k{\rm p1s}}$ times the
probability, $p^{j, k\rm p}_{\rm d}$, to then make the transition
$k{\rm p}\rightarrow j$ (first term).
Using Eq.~\eqref{eq:mod_pd_Dnem_j_c} and \eqref{eq:mod_pd_Dnem_j_e}
leads to Eq.~\eqref{eq:dNjdt_1p1_b}.

Equation~\eqref{eq:dNjdt_1p1_b} helps make the connection of the full
two-photon net rate and the '$1+1$' photon terms because
Eq.~\eqref{eq:dNjdt_1p1_b} can be directly derived from
Eq.~\eqref{eq:dNjdt_2g}, assuming that the two-photon profile is given
by independent (non-interacting) resonances, where the line shapes are
given by the normal Voigt-profiles.

Substituting $\bar{n}_{k{\rm p1s}}=\nbb(\nu_{k{\rm
    p1s}})+\Delta\bar{n}_{k{\rm p1s}}$, and using the relation
$e^{h\nu/k\Tg}=[1+\nbb(\nu)]/\nbb(\nu)$, Eq.~\eqref{eq:dNjdt_1p1_b}
simplifies to
\beal
\label{eq:dNjdt_1p1_simp}
\left.\Abl{X_{j}}{t}\right|^{2\gamma (1+1)}_{{\rm 1s},k{\rm p}}
&= A^\ast_{j\,k{\rm p}} \, p^{k{\rm p}}_{\rm em}
\left[
\frac{g_j}{g_{\rm 1s}}X_{\rm 1s}\,e^{-h\nu_{j1}/k\Tg}-X_{j}
\right]
\nonumber\\
&\qquad\qquad
+ \frac{g_j}{g_{\rm 1s}}X_{\rm 1s}\,A_{j\,k{\rm p}} \, p^{k{\rm p}}_{\rm em}\,
\nbb(\nu_{jk\rm p})\,\Delta \bar{n}_{k{\rm p1s}},
\end{align}
such that upon summing over the intermediate $k$p resonances we have,
\beal
\label{eq:dNjdt_1p1_simp_tot}
\left.\Abl{X_{j}}{t}\right|^{2\gamma (1+1)}_{{\rm 1s}}
&= A^{2\gamma(1+1), \ast}_{j{\rm 1s}}
\left[
\frac{g_j}{g_{\rm 1s}}X_{\rm 1s}\,e^{-h\nu_{j1}/k\Tg}-X_{j}
\right]
\nonumber\\
&\qquad\qquad
+ \frac{g_j}{g_{\rm 1s}}X_{\rm 1s}\sum_{k=2}^{k<n_j} 
A_{j\,k{\rm p}} \, p^{k{\rm p}}_{\rm em}\,\nbb(\nu_{jk\rm p})\,\Delta \bar{n}_{k{\rm p1s}}. 
\end{align}
The above with Eq.~\eqref{eq:dNjdt_2g_tilde} reveals the correction
term for the rate equations as;
\beal
\label{eq:DR2gamma}
\Delta R^{2\gamma}_{j\leftrightarrow {\rm 1s}}
&= 
A^{2\gamma}_{j\rm 1s}\,\Delta \mathcal{G}^{j}_2\,\left[ \frac{g_j}{g_{\rm 1s}}\,X_{\rm 1s} e^{-h\nu_{j\rm 1s}/k\Tg} - X_{j} \right] 
\nonumber\\
&\;
+ A^{2\gamma}_{j{\rm 1s}}\,\frac{g_j}{g_{\rm 1s}}X_{\rm 1s}
\left[
\Delta \mathcal{G}^{j}_1 - 
\sum_{k=2}^{k<n_j} 
\frac{A_{j\,k{\rm p}} \, p^{k{\rm p}}_{\rm em}}{A^{2\gamma}_{j{\rm 1s}}}\,\nbb(\nu_{jk\rm p})\,\Delta \bar{n}_{k{\rm p1s}} 
\right].
\end{align}
Here we define $\Delta \mathcal{G}^{j}_2=\mathcal{G}^{j}_2-
A^{2\gamma(1+1), \ast}_{j{\rm 1s}}/A^{2\gamma}_{j{\rm 1s}}$.

Similar to the 2s-1s two-photon channel, the correction to the rate
equations here has two contributions. The first is related to $\Delta
\mathcal{G}^{j}_2$, which is independent of the solution to the photon
distribution and therefore can be pre-calculated, and the second
arising from the integral $\Delta \mathcal{G}^{j}_1$.  However, in
contrast to the 2s-1s two-photon channel, in the normal rate
equations, part of the latter term is already included. Thus the
'$1+1$' photon term has to be subtracted (last term in brackets), \changeI{where}
this term is calculated using the Sobolev approximation for $\Delta
\bar{n}_{k{\rm p1s}}$.

\changeI{$\Delta
\mathcal{G}^{j}_2$ in principle also arises in the normal '$1+1$' picture, when differentially accounting for the effect of stimulated emission in the CMB blackbody. However, the shape of the two-photon emission profile is crucial, since with the normal sum of Lorentzians the integrand in Eq.~\eqref{eq:2saverage_c} would diverge for $\nu\rightarrow \nu_{j1}$ and $\nu\rightarrow 0$ \citep{Chluba2009}.
Furthermore, the latter two terms in Eq.~\eqref{eq:DR2gamma} account for both, modifications in the shape of the full two-photon profiles, and differences in the solution of the photon field in comparison with the standard Sobolev approximation.
}

The problem is numerical because two large terms are being subtracted.
One way to achieve stable results is to split the range of
integration into intervals where the mean occupation number in the
standard Sobolev approximation is represented by
\citep[compare][]{Chluba2008b}
\beal
\label{eq:Dn_Sobole_x}
\Delta \bar{n}_{k{\rm p1s}}
&= \Delta n^{k \rm p}_{\rm L}
\int_0^\infty \varphi^{k\rm p}_{\rm V} (\nu')\left[1- e^{-\tau^{k\rm p}_{\rm S} [1-\chi^{k\rm p}_{\nu'}]}\right] \id\nu'
\nonumber\\
&= \Delta n^{k \rm p}_{\rm L} \left[
\chi^{k\rm p}_{\nu'}- \frac{e^{-\tau^{k\rm p}_{\rm S} [1-\chi^{k\rm p}_{\nu'}]} }{\tau^{k\rm p}_{\rm S}} 
\right]_0^\infty
=  \Delta n^{k \rm p}_{\rm L} \left[1-P_{\rm S}^{k\rm p} \right],
\end{align}
with $\chi^{k\rm p}_\nu=\int^\nu_0 \varphi^{k\rm p}_{\rm V} \id\nu'$.
Outside the resonances one can simply compute each term in
Eq.~\eqref{eq:DR2gamma} separately, since there the contributions are
small. For those intervals containing a resonance $k$ on the other
hand, one should compute both contributions in one integral, so that
the main terms cancel.
Clearly, the choice of the intervals is only motivated by the
numerical precision that {needs} to be achieved. Since the Voigt-profiles
have their main support inside the Doppler core, it is sufficient to
define regions of a few Doppler width around the resonances. This
approach suffices for our purpose.

Alternatively, one can directly integrate the net two-photon
production rate, Eq.~\eqref{eq:dDn_xdt_j_emabs_tilde}, over frequency and then
subtract the net '$1+1$' photon rate to obtain the correction. We
confirmed that both approaches lead to the same answer.

To capture part of the dependence of $\Delta
R^{2\gamma}_{j\leftrightarrow {\rm 1s}}$ on the solution for the
populations, {in numerical computations} we tabulate the function
$\mathcal{F}^{2\gamma}_{j\leftrightarrow {\rm 1s}}=\Delta
R^{2\gamma}_{j\leftrightarrow {\rm 1s}}/(X_{\rm 1s} \Delta n_{\rm
  L}^{j})$ as a function of redshift, once we computed the solution for the
photon field using the results for the populations of the levels
obtained from a run of our effective multi-level recombination code.

\subsection{Raman-scattering}
\label{sec:Raman-term}
In our previous works \citep{Chluba2009, Chluba2009b} we did not
consider the effect of {\it Raman-scattering} on the ionization
history. However, correction due to this process reaches
$\Delta N_{\rm e}/N_{\rm e} \sim 0.9\%$ at $z\sim 900$
\citep{Hirata2008}, and hence demands careful consideration.
The matrix element for this process is directly related to the one for
the two-photon emission process by crossing-symmetry. In
Appendix~\ref{app:comp_two_photon_profs} we explain how to compute the
Raman-scattering profiles, $\varphi^{\rm R}_{j} (\nu)$, for the $j$-1s
Raman process.
Additional details can also be found in \citet{Hirata2008}, where
the importance of this effect during recombination was shown for the
first time.

\subsubsection{Net rates for $n$s-1s and $n$d-1s Raman-scattering}
The net change in the number density of electrons in the level
$j~\in~\{n{\rm s}, n{\rm d}\}$ caused by $j$-1s Raman scatterings is
given by
\bsub
\label{eq:dNjdt_R}
\beal
\label{eq:dNjdt_R_a}
\left.\Abl{X_{j}}{t}\right|^{\rm R}_{\rm 1s}
&=\frac{g_j}{g_{\rm 1s}}A^{\rm R}_{j\rm 1s}X_{\rm 1s}\int_{\nu_{j\rm 1s}}^{\nu_{\rm 1s c}}
\varphi^{\rm R}_{j} (\nu-\nu_{j\rm 1s})\, n(\nu)\,[1+n(\nu-\nu_{j\rm 1s})]\id\nu
\nonumber
\\
&\qquad
-A^{\rm R}_{j\rm 1s}X_{j}\int_0^{\nu_{j\rm c}} \varphi^{\rm R}_{j} (\nu) \,n(\nu)\,[1+n(\nu_{j\rm 1s}+\nu)]\id\nu
\\
\label{eq:dNjdt_R_b}
&\equiv \frac{g_j}{g_{\rm 1s}}A^{\rm R}_{j\rm 1s}X_{\rm 1s}\int_{\nu_{j\rm 1s}}^{\nu_{\rm 1s c}}
\varphi^{\rm R}_{j} (\nu-\nu_{j\rm 1s})\, n(\nu)\,[1+n(\nu-\nu_{j\rm 1s})]\id\nu
\nonumber
\\
&
-A^{\rm R}_{j\rm 1s}X_{j} \int_{\nu_{j\rm 1s}}^{\nu_{\rm 1s c}} 
\varphi^{\rm R}_{j} (\nu-\nu_{j\rm 1s}) \,n(\nu-\nu_{j\rm 1s})\,[1+n(\nu)]\id\nu,
\end{align}
\esub
where Eq.~\eqref{eq:dNjdt_R_b} was simply obtained from Eq.~\eqref{eq:dNjdt_R_a} by transforming the frequency range of the second integral.
In Eq.~\eqref{eq:dNjdt_R}, $\varphi^{\rm R}_{j}$ denotes the $j$-1s Raman-scattering profile, and the Raman-scattering coefficient is given by\footnote{We call $A^{\rm R}_{j\rm 1s}$ 'coefficient' since in vacuum there is no Raman-process.}
\beal
\label{eq:AR_j1s}
A^{\rm R}_{j\rm 1s}&=\sum_{k=n_j+1}^{n_{\rm max}} \frac{g_{k{\rm p}}}{g_j}\,A_{k{\rm p}\, j} \, p^{k{\rm p}}_{\rm em}.
\end{align}
The ratio of the statistical weights is $g_{k\rm p}/g_{j}=3$ for the $n$s-states, and $g_{k\rm p}/g_{j}=3/5$ for $n$d-states.
Equation~\eqref{eq:AR_j1s} simply reflects the one-photon terms and branching ratios of all the '$1+1$' photon routes $j\rightarrow n{\rm p}\rightarrow {\rm 1s}$ via intermediate p-states with $n>n_j$. 

A rigorous treatment of Eq.~\eqref{eq:AR_j1s} would include the
integral over continuum states. However, any electron reaching the
continuum would forget its history because of fast Coulomb
interactions resulting in decoherence of the Raman process in the continuum.
Furthermore, as mentioned above, the Lyman-continuum is extremely
optically thick such that these channels will always cancel out
\citep[see also][]{Hirata2008}.
Also, in numerical computations we only follow the evolution of the
photon field up to some maximal frequency, $\nu_{\rm max}$. Therefore,
in our description we are not accounting for the full Raman-process
connected with transitions involving photons with $\nu >\nu_{\rm
  max}$. This approximation is fully justified as the higher
Lyman-series contribute negligible amounts to the total recombination
rate. Thus the sum over intermediate p-states {become} finite, {without significant loss of precision}.

To simplify Eq.~\eqref{eq:dNjdt_R} we define the following quantities\footnote{Formally, the
  upper limit of the integral over the Raman-profiles should go to
  infinity. However, since we are following the spectrum in a finite
  range of frequencies, this introduces an upper limit, $\nu_{\rm
    max}\leq \nu_{\rm 1s c}$.}
\bsub
\label{eq:Raverage}
\beal
\label{eq:Raverage_a}
\twosavg{f(\nu)}{\rm R}{i}&=\int^{\nu_{\rm max}}_{\nu_{i\rm 1s}} \varphi^{\rm R}_{i} (\nu-\nu_{j\rm 1s})\, f(\nu) \id\nu
\\
\mathcal{R}^{i}_1&=\twosavg{n\,[1+n']}{\rm R}{i}
\\
\mathcal{R}^{i}_2&=\twosavg{n'[1+n]}{\rm R}{i}
\end{align}
\esub
with $n'=n(\nu-\nu_{j\rm 1s})$.
In the spirit of the two-photon emission and absorption process, we can now write
\bsub
\label{eq:Raverage_approx}
\beal
\mathcal{R}^{i}_1&\approx \twosavg{\nbb\,[1+\nbb']}{\rm R}{i} + \twosavg{\Delta n\,[1+\nbb']}{\rm R}{i}
\nonumber\\
&\approx \mathcal{R}^{i}_2 \, e^{-h\nu_{j\rm 1s}/k \Tg}+ \twosavg{\Delta n\,[1+\nbb']}{\rm R}{i}
\\
\mathcal{R}^{i}_2& \approx \twosavg{\nbb'[1+\nbb]}{\rm R}{i}
= A_{j \rm 1s}^{\rm R, \ast}/A_{j \rm 1s}^{\rm R}.
\end{align}
\esub
The total $j\rightarrow \rm 1s$  '$1+1$'  Raman-scattering rate in the CMB blackbody ambient radiation field is
\beal
\label{eq:A_R_stim}
A_{j \rm 1s}^{\rm R(1+1), \ast}= \sum_{k=n_j+1}^{n_{\rm max}}
\frac{g_{k\rm p}}{g_{j}} A_{k{\rm p}\,j} \,  \nbb(\nu_{jk})\, p^{k{\rm p}}_{\rm em}.
\end{align}
This then leads to 
\bsub
\label{eq:dNjdt_R_tilde}
\beal
\label{eq:dNjdt_R_tilde_a}
\left.\Abl{N_{j}}{t}\right|^{\rm R}_{\rm 1s}
&=A^{\rm R, \ast}_{j\rm 1s}\left[ \frac{g_j}{g_{\rm 1s}}\,X_{\rm 1s} e^{-h\nu_{j\rm 1s}/k\Tg} - X_{j} \right]
+ \frac{g_j}{g_{\rm 1s}}\, A^{\rm R}_{j \rm 1s}\,X_{\rm 1s}\,\Delta \mathcal{R}^{j}_1
\\[1mm]
\label{eq:dNjdt_R_tilde_b}
\Delta \mathcal{R}^{j}_1
&=
\int^{\nu_{\rm max}}_{\nu_{j\rm1s}}
\varphi^{\rm R}_{j}(\nu-\nu_{j \rm 1s}) \,{[1+n^{\rm pl}}'] \,\Delta n(\nu) \id\nu.
\end{align}
\esub
Here $A^{\rm R, \ast}_{j\rm 1s}$ and $\Delta \mathcal{R}^{j}_1$ are
important in defining the correction to the rate equations (see
\S~\ref{sec:corr_Raman}).
\changeI{Again $A^{\rm R, \ast}_{j\rm 1s}$ is the thermal contribution, while $\Delta \mathcal{R}^{j}_1$ arises from non-thermal photons.}

\subsubsection{Terms in the radiative transfer equation for Raman-scattering}
From Eq.~\eqref{eq:dNjdt_R} the terms in the
radiative transfer equation for the photon field can be obtained.
However, one aspect is important to keep in mind: a photon that
Raman-scatters off an electron in the $j^{\rm th}$-state is removed
from frequencies $0\leq \nu\leq \nu_{j\rm c}$. However, the scattered
photon appears in the frequency range $\nu_{j\rm c}< \nu_{j1}\leq \nu'
\leq \nu_{1\rm c}$, and likewise for the inverse process.
This description assumes {\it complete redistribution} of photons over the full Raman-scattering profile\footnote{We neglect corrections caused by partial redistribution in Raman-scattering events, but like in the case of two-photon transitions these should be very small.} during each scattering event.
Therefore, the terms for the radiative transfer equation read
\bsub
\label{eq:dNnu_jdt_R}
\beal
\label{eq:dNnu_jdt_R_a}
\left.\frac{1}{c}\,\frac{\partial N_{\nu}}{\partial t}\right|^{j\rm 1s}_{{\rm R}, \nu\leq  \nu_{j\rm c}}
&=
\frac{g_j}{g_{\rm 1s}}A^{\rm R}_{j\rm 1s}N_{\rm 1s}\,
\tilde{\varphi}^{\rm R}_{j} (\nu)\, n(\nu_{j\rm 1s}+\nu)\,[1+n(\nu)]
\nonumber
\\
&\qquad\quad
-A^{\rm R}_{j\rm 1s}N_{j} \,\tilde{\varphi}^{\rm R}_{j} (\nu) \,n(\nu)\,[1+n(\nu_{j\rm 1s}+\nu)]
\\[1mm]
\label{eq:dNnu_jdt_R_b}
\left.\frac{1}{c}\,\frac{\partial N_{\nu}}{\partial t}\right|^{j\rm 1s}_{{\rm R}, \nu_{j{\rm 1s}}\leq \nu}
&=
A^{\rm R}_{j\rm 1s}N_{j} \,\tilde{\varphi}^{\rm R}_{j} (\nu-\nu_{j\rm 1s}) \,n(\nu-\nu_{j\rm 1s})\,[1+n(\nu)]
\nonumber
\\
&\!\!\!\!\!\!\!
-\frac{g_j}{g_{\rm 1s}}A^{\rm R}_{j\rm 1s}N_{\rm 1s}\,
\tilde{\varphi}^{\rm R}_{j} (\nu-\nu_{j\rm 1s})\, n(\nu)\,[1+n(\nu-\nu_{j\rm 1s})],
\end{align}
\esub
where $\tilde{\varphi}^{\rm R}_{j} (\nu)= \varphi ^{\rm R}_{j} (\nu)/4\pi$. 
It is clear that the total integral over frequency vanishes, when
adding the above two terms, showing that the Raman-process conserves
photon number. However, the number density of electrons in the 1s and
$j$-state is altered after each Raman-scattering event, according to
Eq.~\eqref{eq:dNjdt_R}.

With regards to the recombination dynamics we are not interested in the
changes to the photon spectrum at low frequencies. Therefore, we only
consider Eq.~\eqref{eq:dNnu_jdt_R_b}.  For stimulated terms,
the distortions can be neglected. Furthermore, one can define
$\tilde{\varphi}^{\rm R, \ast}_{j} (\nu)\equiv \tilde{\varphi}^{\rm
  R}_{j} (\nu)\, \nbb(\nu)\, [1+\nbb(\nu_{j\rm 1s}+\nu)]\approx
\tilde{\varphi}^{\rm R}_{j} (\nu)\, \nbb(\nu)$, and neglect the
distortions at low frequencies, such that
\beal
\label{eq:dNnu_jdt_R_fin}
\left.\frac{1}{c}\,\frac{\partial N_{\nu}}{\partial t}\right|^{j\rm 1s}_{{\rm R}, \nu_{j{\rm 1s}}\leq \nu}
&\approx
A^{\rm R}_{j\rm 1s}N_{j} \,\tilde{\varphi}^{\rm R, \ast}_{j} (\nu-\nu_{j\rm 1s}) 
\nonumber
\\
&\quad\quad
-\frac{g_j}{g_{\rm 1s}}A^{\rm R}_{j\rm 1s}N_{\rm 1s}\, \tilde{\varphi}^{\rm R, \ast}_{j} (\nu-\nu_{j\rm 1s})\,f^{j}_\nu\, n(\nu),
\end{align}
where $f^{j}_\nu$ is defined by Eq.~\eqref{eq:Dnem_j_b}.
In terms of the photon occupation number this equation becomes,
\beal
\label{eq:dDn_xdt_j_R_tilde}
\left.\frac{\partial \Delta n_x}{\partial z}\right|^{j\rm 1s}_{{\rm R}, \nu_{j{\rm 1s}}\leq \nu}
&=
- \frac{\sigma^{\rm R}_{j\rm 1s} N_{\rm 1s} c}{H[1+z]}\,\frac{\nu^2_{j1}}{\nu^2_{21}}\,\frac{\phi^{\rm R,\ast}_{j}}{x^2}
\left[\nu_{\rm 21}\,\Delta n_{\rm em}^{j} - f^{j}_x \Delta n_x \right],
\end{align}
where $\Delta n_{\rm em}^{j}$ is defined as in Eq.~\eqref{eq:Dnem_j_a}.
The $j$-1s Raman-scattering cross section is given by
$\sigma^{\rm R}_{j \rm 1s}=\frac{g_j}{g_{\rm
    1s}}\,\frac{\lambda^2_{j1} A^{\rm R}_{j\rm 1s}}{8\pi \nu_{j1}}$,
and we set $\phi^{\rm R, \ast}_{j}\equiv
4\,\pi\,\nu_{j1}\tilde{\varphi}^{\rm R,\ast}_{j}(\nu-\nu_{j\rm 1s})$.
Note the close similarity of this equation to the one-photon equation
for the Lyman-series emission and absorption channels in
Eq.~\eqref{eq:Line_em_ab_k}.
Photons scattering from frequencies $0\leq \nu\leq \nu_{j\rm c}$ into
the range $\nu_{j\rm 1s}\leq \nu$ appear as a source term. This is
related to the fact the these photons are drawn from the CMB
blackbody.

\subsubsection{Correcting the Lyman-series emission and absorption terms in the radiative transfer equation}
\label{sec:double_3}
Like in the case of two-photon emission and absorption, the resonant
part of the Raman-process is already part of the '$1+1$' photon
Lyman-series transfer in Eq.~\eqref{eq:Line_em_ab_k}. 
To avoid double-counting we simply {have} to correct the death probability and $\Delta n_{\rm em}^{k\rm p}$
of the Lyman-k resonance for terms that are included in the
Raman-scattering process.
For example, when using the terms for the 2s-1s Raman-scattering
process in the radiative transfer equation, $p^{k\rm p}_{\rm d}$ and
$\Delta n_{\rm em}^{k\rm p}$ for Lyman-$\beta$, $\gamma$, $\delta$,
and higher will have to be corrected.

The modified death probability can be obtained by adding appropriate
terms to the sums of Eq.~\eqref{eq:mod_pd_Dnem_j_all}.
However, for each included {Raman-channels} one now has $R^{j, -}_{k\rm
p}=A_{k{\rm p}\,j}\,[1+\nbb(\nu_{j\, k{\rm p}})]$, and $R^{j, +}_{k\rm
  p}=\frac{g_{k\rm p}}{g_j}\,A_{k{\rm p}\,j}\,\nbb(\nu_{j\, k{\rm
    p}})\,X_j\equiv R^{j, -}_{k\rm p}\,\frac{g_{k\rm
    p}}{g_j}\,X_j\,e^{-h\nu_{j\, k{\rm p}}/k\Tg}$.

\subsubsection{Correcting the net rates in the multi-level atom}
\label{sec:double_4}
\label{sec:corr_Raman}
Like in the case of two-photon emission and absorption events,
corrections to the net rates in the multi-level atom have to be
defined to avoid double-counting.
In the standard multi-level description of all $j\leftrightarrow
n{\rm p}\leftrightarrow {\rm 1s}$ sequences ($n_j<n$), the
contributions to the Raman-scattering net rate, Eq.~\eqref{eq:dNjdt_R},
takes the form
\bsub
\label{eq:dNjdt_1p1_R}
\beal
\label{eq:dNjdt_1p1_R_a}
\left.\Abl{X_{j}}{t}\right|^{\rm R (1+1)}_{{\rm 1s},k{\rm p}}
&=\!\!
\frac{g_{k\rm p}}{g_{\rm 1s}} X_{\rm 1s}\,A_{k{\rm p1s}} \, p^{j, k\rm p}_{\rm d}\,\bar{n}_{k{\rm p1s}}
\!-\!\frac{g_{k\rm p}}{g_{j}} X_{j}\, A_{k{\rm p} j} \, \nbb(\nu_{jk})  p^{k{\rm p}}_{\rm em}
\\
\label{eq:dNjdt_1p1_R_b}
&\equiv\!
\frac{g_{k\rm p}}{g_{\rm 1s}}X_{\rm 1s}\,A^\ast_{k{\rm p} j} \, p^{k{\rm p}}_{\rm em}\,\bar{n}_{k{\rm p1s}}
-\frac{g_{k\rm p}}{g_{j}} X_{j}\, A_{k{\rm p} j} \, \nbb(\nu_{jk}) p^{k{\rm p}}_{\rm em}
\\
\label{eq:dNjdt_1p1_R_c}
\left.\Abl{X_{j}}{t}\right|^{\rm R (1+1)}_{\rm 1s}
&=
\sum_{k=n_j+1}^{n_{\rm max}} \left.\Abl{X_{j}}{t}\right|^{\rm R (1+1)}_{{\rm 1s},k{\rm p}}.
\end{align}
\esub
As mentioned above, $n_{\rm max}$ is a consequence of the finite
computational domain.
The terms in Eq.~\eqref{eq:dNjdt_1p1_R} is interpreted as 
in the case of two-photon emission and absorption (see
\S~\ref{sec:corr_2g}).

Inserting $\bar{n}_{k{\rm p1s}}=\nbb(\nu_{k{\rm
    p1s}})+\Delta\bar{n}_{k{\rm p1s}}$, and using the relation
$e^{h\nu/k\Tg}=[1+\nbb(\nu)]/\nbb(\nu)$, Eq.~\eqref{eq:dNjdt_1p1_R_b} simplifies to,
\beal
\label{eq:dNjdt_1p1_R_simp}
\left.\Abl{X_{j}}{t}\right|^{\rm R (1+1)}_{{\rm 1s},k{\rm p}}
&= \frac{g_{k\rm p}}{g_{j}} A_{k{\rm p}\,j} \,  \nbb(\nu_{jk}) p^{k{\rm p}}_{\rm em}
\left[
\frac{g_j}{g_{\rm 1s}}X_{\rm 1s}\,e^{-h\nu_{j1}/k\Tg}-X_{j}
\right]
\nonumber\\
&\qquad\qquad
+ \frac{g_{k\rm p}}{g_{\rm 1s}}X_{\rm 1s}\,A^\ast_{k{\rm p}\,j} \, p^{k{\rm p}}_{\rm em} \, \Delta \bar{n}_{k{\rm p1s}}, 
\end{align}
such that summing over the intermediate $k$p resonances leads to,
\beal
\label{eq:dNjdt_1p1_simp_tot}
\left.\Abl{X_{j}}{t}\right|^{\rm R (1+1)}_{{\rm 1s}}
&= A^{\rm R, \ast}_{j{\rm 1s}}
\left[
\frac{g_j}{g_{\rm 1s}}X_{\rm 1s}\,e^{-h\nu_{j1}/k\Tg}-X_{j}
\right]
\nonumber\\
&\qquad\qquad
+ X_{\rm 1s}\sum_{k=n_j+1}^{n_{\rm max}} 
\frac{g_{k\rm p}}{g_{\rm 1s}}\,A^\ast_{k{\rm p}\,j} \, p^{k{\rm p}}_{\rm em}\,\Delta \bar{n}_{k{\rm p1s}}.
\end{align}
Using Eq.~\eqref{eq:dNjdt_R_tilde} it is clear that the correction term for the rate equations are,
\beal
\label{eq:DR_R}
\Delta R^{\rm R}_{j\leftrightarrow {\rm 1s}}
&= A^{\rm R}_{j\rm 1s}\,\Delta \mathcal{R}^{j}_2\,\left[ \frac{g_j}{g_{\rm 1s}}\,X_{\rm 1s} e^{-h\nu_{j\rm 1s}/k\Tg} - X_{j} \right] 
\nonumber\\
&\;
+ A^{\rm R}_{j{\rm 1s}}\,\frac{g_j}{g_{\rm 1s}}X_{\rm 1s}
\left[
\Delta \mathcal{R}^{j}_1 - 
\sum_{k=n_j+1}^{n_{\rm max}} 
\frac{g_{k\rm p}}{g_{j}}\,
\frac{A^\ast_{k{\rm p}\,j} \, p^{k{\rm p}}_{\rm em}}{A^{\rm R}_{j{\rm 1s}}}\,\Delta \bar{n}_{k{\rm p1s}} 
\right],
\end{align}
where we define $\Delta \mathcal{R}^{j}_2=\mathcal{R}^{j}_2- A^{\rm R(1+1), \ast}_{j{\rm 1s}}/A^{\rm R}_{j{\rm 1s}}$.
\changeI{Like for the two-photon channels (see Eq.~\eqref{eq:DR2gamma}), $\Delta
\mathcal{R}^{j}_2$ in principle also arises in the normal '$1+1$' picture, where the shape of the Raman-profile ensure that the integrand remains finite, this time in the limit of $\nu\rightarrow \nu_{j\rm 1s}$ (see Appendix~\ref{app:Raman_Profiles}).
Furthermore, the latter two terms in Eq.~\eqref{eq:DR_R} account for both, modifications in the shape of the Raman profiles with respect to the normal sum of Lorentzians, and differences in the solution of the photon field with respect to the Sobolev approximation.
}

Again one can compute the integrals over frequency by splitting the range of integration and using Eq.~\eqref{eq:Dn_Sobole_x} to model the ``Sobolev part''.
In numerical calculations, we tabulate $\mathcal{F}^{\rm
  R}_{j\leftrightarrow {\rm 1s}}=\Delta R^{\rm R}_{j\leftrightarrow
  {\rm 1s}}/(X_{\rm 1s}\,\Delta n^j_{\rm L})$ versus redshift, to
include the correction into the effective multi-level recombination
code, and then use this to correct the rate equations.

\begin{figure*}
\centering
\includegraphics[width=0.95\columnwidth]{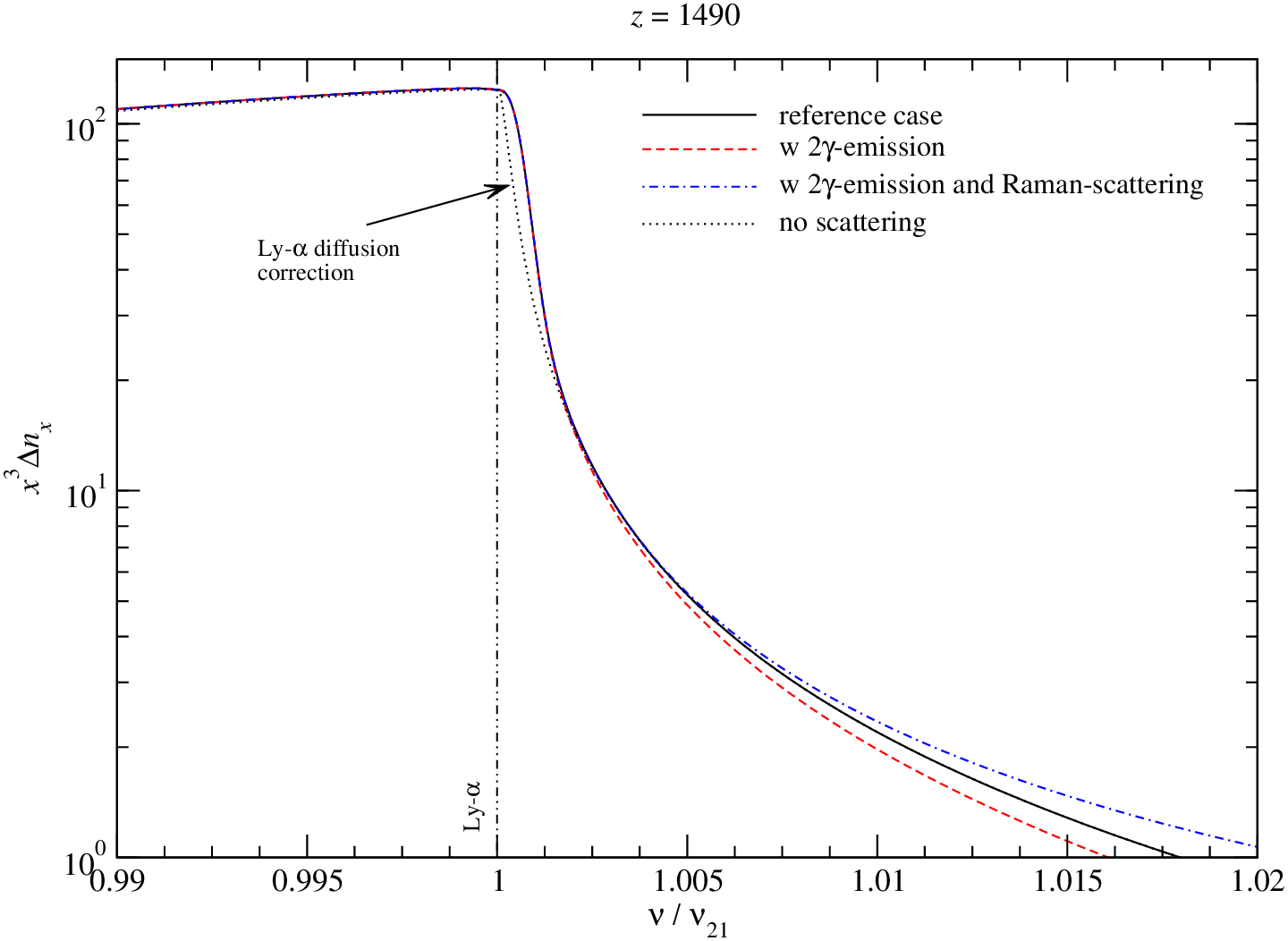}
\hspace{4mm}
\includegraphics[width=0.95\columnwidth]{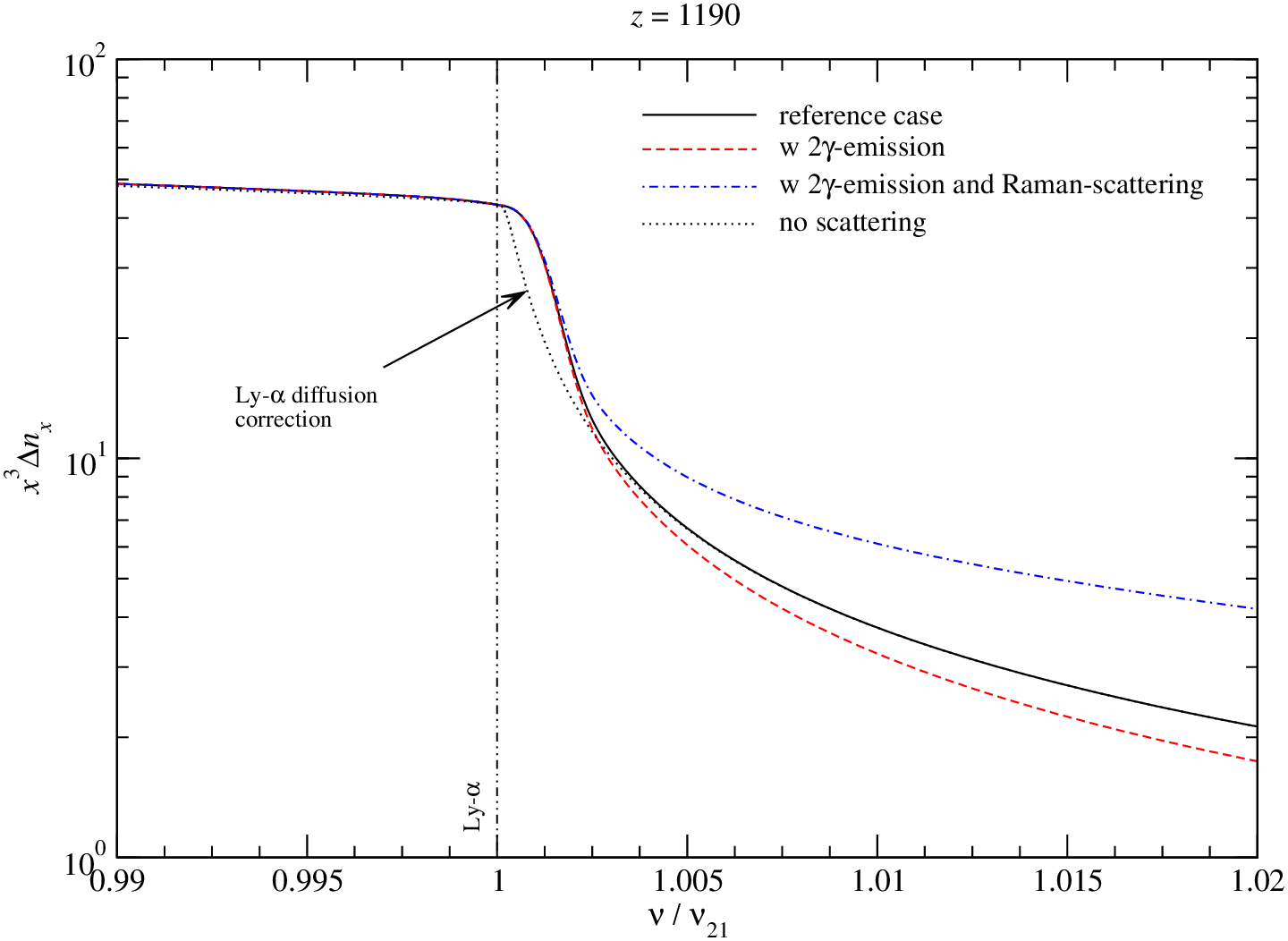}
\\[2mm]
\includegraphics[width=0.95\columnwidth]{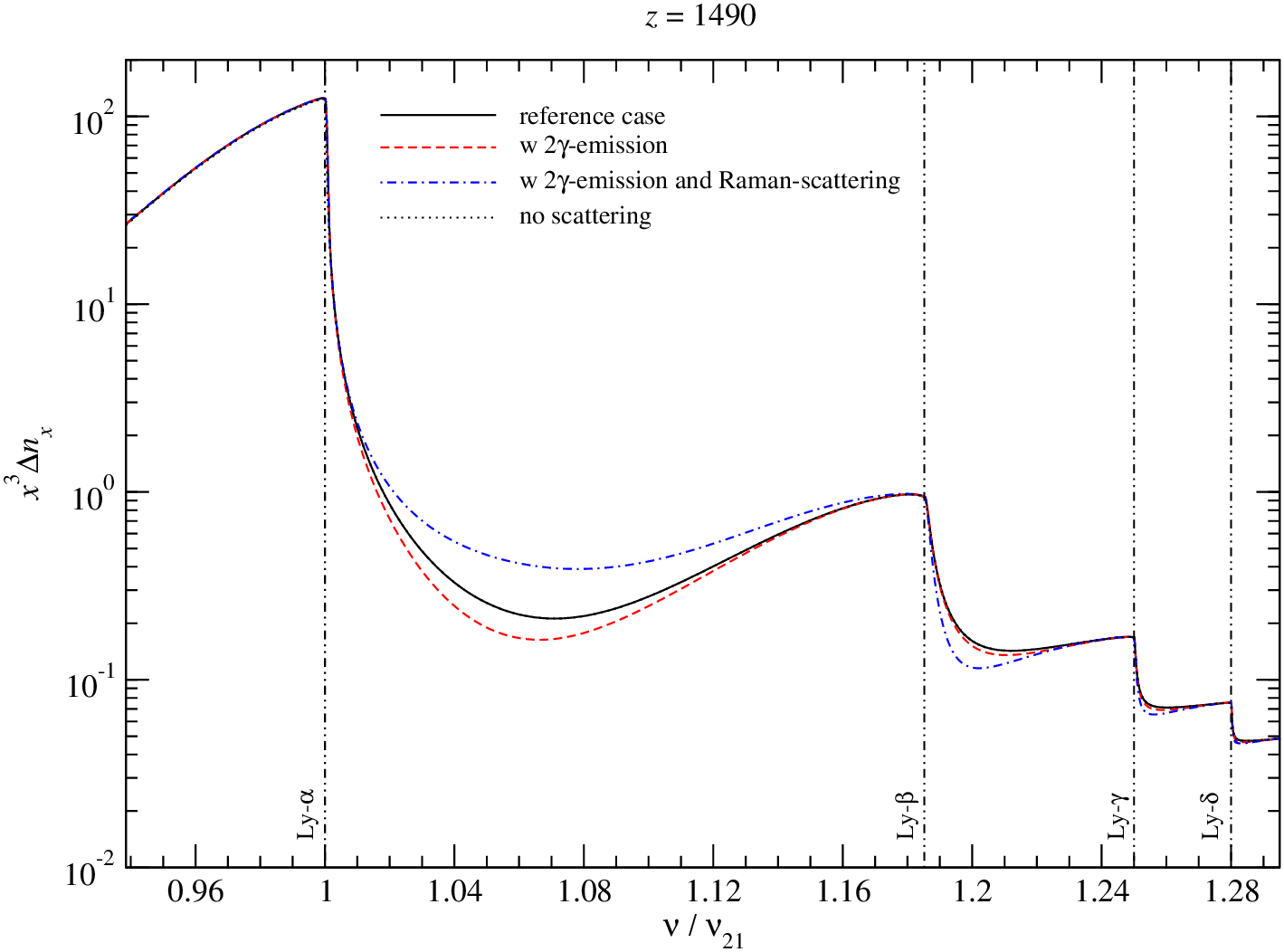}
\hspace{4mm}
\includegraphics[width=0.95\columnwidth]{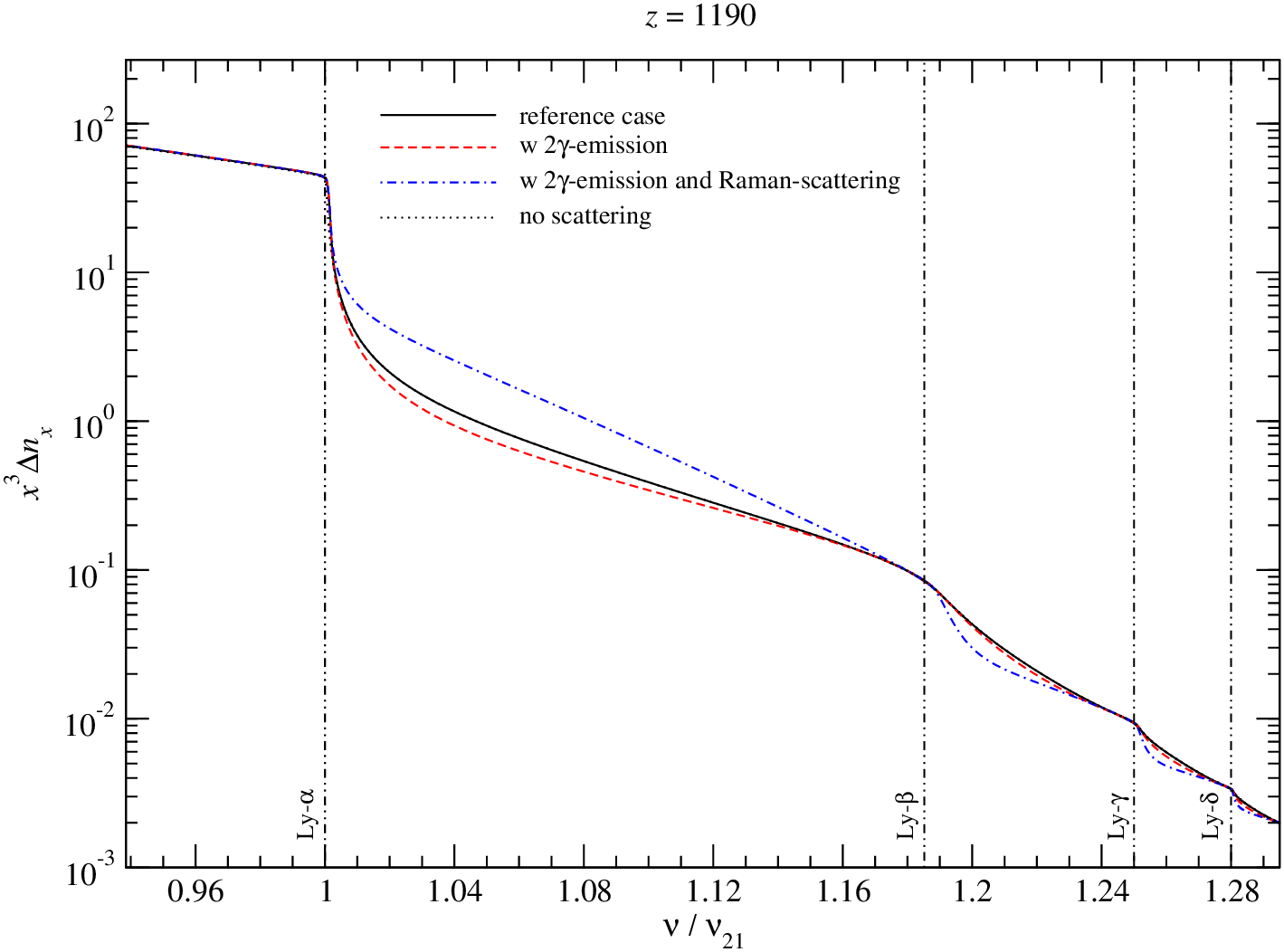}
\caption{Solution for the Lyman-series distortion at $z=1490$ (left
  panels) and $z=1190$ (right panels) for different combinations of
  physical processes (for details see \S\ref{sec:DI-Ly-n_cases}).
  In all cases we include the effect of electron scattering. We also
  marked the positions of the Lyman-series resonances with the
  vertical dashed-dot-dotted lines. The effect of partial frequency
  redistribution is only important close to the Lyman-$\alpha$ line
  center, so that the dotted line is only visible in the upper panels.
  \changeI{A movie on the time-evolution of the Lyman-series distortion can be found at www.Chluba.de/Lyman-series-movie.}
}
\label{fig:DI.Ly-series.n_5}
\end{figure*}

\section{Changes in the Lyman-series distortion for different physical processes}
\label{sec:DI-Ly-n_cases}
In \S\ref{sec:DNe_cases} we discuss the changes to the free electron
fraction due to the various physical processes under
consideration. However, in order to understand the source of these
corrections it is illustrative to first look at the modifications in
the Lyman-series spectral distortion.

In Fig.~\ref{fig:DI.Ly-series.n_5} we present the spectral distortion
at two different redshifts, one before the maximum of the Lyman-series
emission (which happens at $z\sim 1300-1400$), and one just before the
maximum of the Thomson visibility function.
We include Lyman-resonances up to $n=8$ for these computations. The
solutions to the populations of the hydrogen levels were obtained
from our implementation of the effective 400-shell recombination
code.
The solid black line in all panels, shows our {\it reference case},
for which the Lyman-series is modelled using Voigt-profiles.
This case already includes the effect of {\it resonance scattering}
(for all Lyman-series resonances), {\it electron scattering}, the full
{\it time-dependence} \citep{Chluba2008b} of the emission and
absorption process, and the {\it thermodynamic correction} factor for
each resonance \citep{Chluba2009}, capturing {a large part} of the
corrections with respect to the {Sobolev treatment}. In
particular, the distinction between scattering and
emission/absorption events (by introducing the death probability) is
important for the photon distribution on the blue side of the
Lyman-$\alpha$ resonance \citep[see discussion in][]{Chluba2008b}.
{Furthermore,} time-dependence and the thermodynamic correction factor
lead to a large modification of the photon distribution with respect
to the standard Sobolev case.

We will now discuss the effect of the different processes on the
shape of the Lyman-series distortion separately.

\subsection{Effect of Lyman-series scattering}
In Fig.~\ref{fig:DI.Ly-series.n_5}, the dotted curve shows the case
for which we ``switched off'' the terms for Lyman-series scattering.
This line is only visible in the upper panels, since at high
frequencies above the Lyman-$\alpha$ line it coincides with the
reference case.
The figure illustrates {that} partial redistribution by Lyman-series
scattering is only important close to the Lyman-$\alpha$ resonance,
and on its red wing.
We could, in principle, neglect the correction due to resonance
scattering for Lyman-$n$ with $n>2$, however, with our efficient PDE
solver it is straightforward to take them into account.

The physical reason for this behaviour is that the scattering
probability in the Lyman-$\alpha$ line is very close to unity ($p_{\rm
  sc}^{\rm 2p}\sim 0.999 - 0.9999$), such that only in the vicinity of
the Doppler core can real emission and absorption terms act
efficiently, strongly redistributing photons over frequency.
Outside the Doppler-core, however, redistribution is much slower making
the effect of Doppler redistribution visible.

For the higher Lyman-series resonance, on the other hand, the death
probability is only about an order of magnitude smaller than the
scattering probability, implying that far out in the wings of the
resonance photons can be efficiently redistributed by emission and
absorption processes. In this case, resonance scattering leads to a
small correction \citep[see also arguments in][]{Yacine2010b}.

\subsection{Two-photon emission and absorption from the excited states with $n\geq 3$}
Next we include the corrections due to the {\it
  shapes} of the $n$s-1s and $n$d-1s two-photon profiles
(Fig.~\ref{fig:DI.Ly-series.n_5}, red/dashed line).
By {\it shape} we also address modifications caused by the presence of CMB
blackbody photons.

One can see that in comparison to the reference case this slightly
decreases the spectral distortion between all Lyman-resonances,
indicating that the emission/absorption opacity has decreased.
The largest effect is seen between the Lyman-$\alpha$ and
Lyman-$\beta$ lines, as a result of the 3s-1s and 3d-1s two-photon
emission and absorption process.
This result is in agreement with our earlier treatment
\citep{Chluba2009}, where it was demonstrated that the shape of the
3s-1s and 3d-1s two-photon profiles leads to a slight acceleration of
recombination, which however, is less important than the corrections
arising from the thermodynamic correction factor and time-dependence,
individually.

We tried to identify the main source of the modifications above the
Lyman-$\beta$ resonance in more detail.
In the full two-photon picture, the 3s-1s and 3d-1s two-photon
emission and absorption channels only act on photons with $\nu\leq
\nu_{31}$. However, when neglecting the modifications to the shapes of
the two-photon profiles, a large part of the opacity above the
Lyman-$\beta$ line (incorrectly) comes from the 3s-1s and 3d-1s
'$1+1$' photon channel, which involves the Lyman-$\alpha$ resonance
and is modelled by a normal Voigt profile.
It turns out that only for the 3s-1s and 3d-1s two-photon process does the
exact shape of the two-photon profile really matter. Above the
Lyman-$\beta$ line the small correction with respect to the solid line
is practically captured by {\it truncating} the Voigt profiles (in
particular the one for Lyman-$\alpha$), {such that the energy is conserved}
(e.g. photons emitted or absorbed in a 3s-1s and 3d-1s
'$1+1$' photon process can only have energies $\nu\leq \nu_{31}$, and
so on).
This illustrates how important the shape of the line profiles is
when going far into the damping wings of the resonances.

\subsection{Importance of Raman scattering}
In addition to the two-photon corrections, we ran cases that also
include the full Raman scattering treatment
(Fig.~\ref{fig:DI.Ly-series.n_5}, blue/dash-dotted line).
One can see that the Raman process led to an enhancement of the
spectral distortion between the Lyman-$\alpha$ and Lyman-$\beta$
resonance, while in all the other cases the spectral distortion
decreases between the resonances.
Thus one expects an increased blue-wing feedback correction and
hence a delay of hydrogen recombination from Lyman-$\alpha$. On the
other hand, these additional red-wing Lyman-$\beta$ photons {were} created in
a 2s-1s Raman event, such that at earlier times an acceleration is
expected. This simple picture is in agreement with earlier discussions
of this process \citep{Hirata2008}.
In the case considered, the main source of the difference in the
Lyman-series distortion comes from the 2s-1s Raman treatment.
Neglecting the Raman-corrections to the higher $n$s-1s and $n$d-1s
channels does no affect the shape of the distortion noticeably.
This is one of the reasons why the Raman process need to be included
only for the first few levels.

Since in the case of Raman scattering, the $2{\rm s}\rightarrow 1{\rm
  s}$ scattering profile is given by $\tilde{\varphi}^{\rm R,
  \ast}_{j} (\nu)\approx \tilde{\varphi}^{\rm R}_{j} (\nu)\,
\nbb(\nu)$ (see Sect.~\ref{sec:Raman-term}), one expects two sources
of corrections: (i) due to the difference of $\varphi^{\rm R}_{j}
(\nu)$ with respect to a sum of Voigt profiles with appropriate
weights, and (ii) the factor $\nbb(\nu)$. In the normal '$1+1$' photon
picture this factor would not appear differentially, but instead
directly for each resonance frequency.
It turns out that both part of the correction are important for the
2s-1s Raman treatment.

We note, that the spectral distortion at $z=1190$ in the full
treatment looks very similar to the curve given in
\citet{Hirata2008}. However, in \citet{Hirata2008} also the CMB
blackbody spectrum was added, and $n_\nu$ instead of $x^3 \Delta n_x$
was plotted, which makes a direct comparison more difficult.

\section{Changes to the free electron fraction for different physical processes}
\label{sec:DNe_cases}
In this section we present our analysis of the different corrections
to the standard recombination calculation. We focus on hydrogen, and
model the helium recombination dynamics using the description given in
\citet{Chluba2009c}, including the first five shells with full
feedback.
With the current version of our effective multi-level recombination
code we are able to account for all important corrections to the
recombination dynamics of hydrogen.
We show a direct comparison with previous results and find very
good agreement.
All figures in which we compared the output of our
recombination code with {\sc Recfast} we used {\sc Recfast v1.4.2}
\citep{Wong2007}, but excluded the corrections to the helium
recombination history in {\sc Recfast} and removed the switches in the
{\sc Recfast} ODE system (see Fendt et al. 2009, for details).
The cumulative effect on the ionization history is presented in
Fig.~\ref{fig:DN_total}.

\subsection{Results from our effective multi-level code}
\label{sec:DNe_EMLC}
\begin{figure}
\centering
\includegraphics[width=\columnwidth]{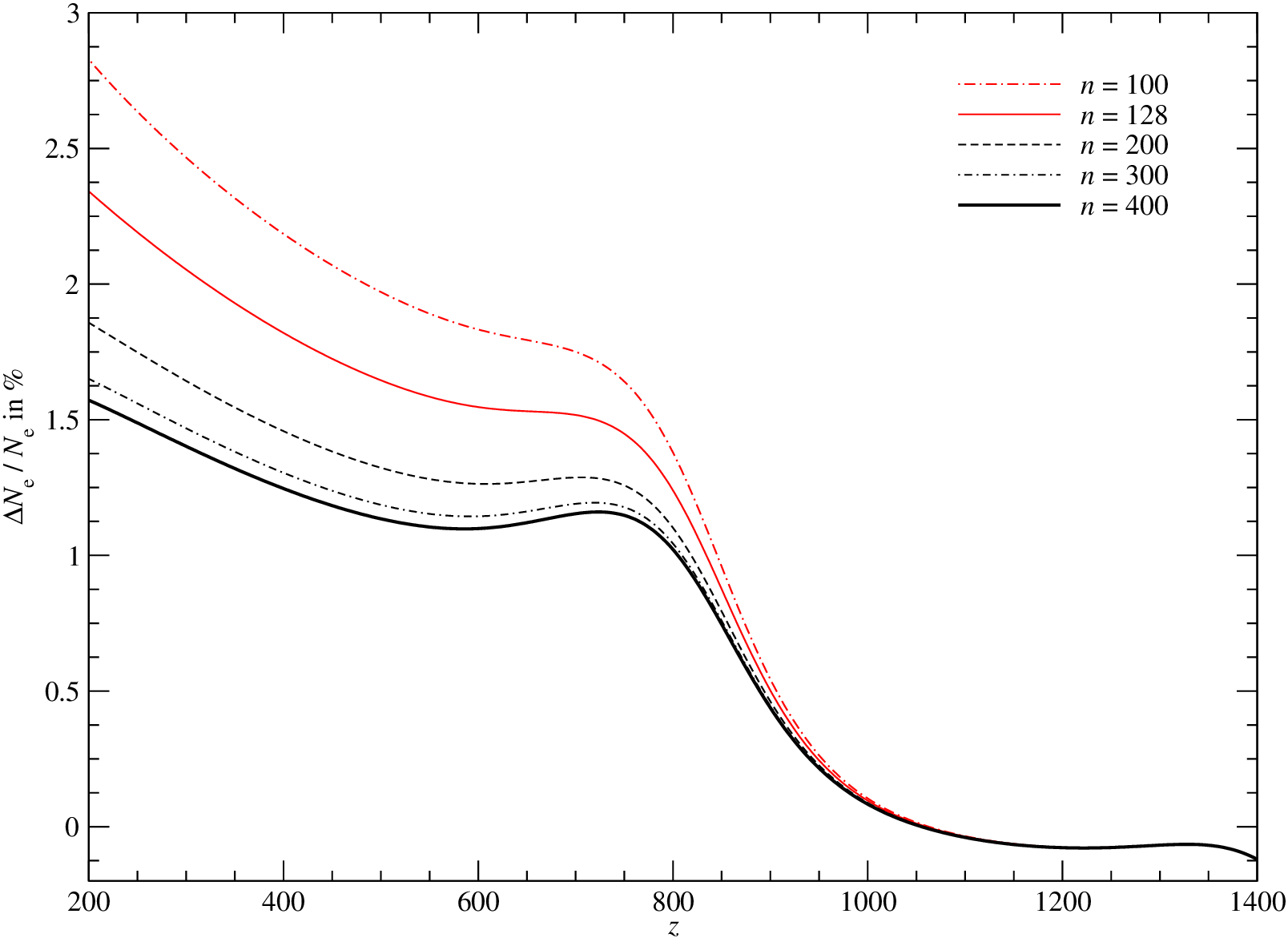}
\caption{Dependence of the modifications to the recombination dynamics
 Êon the number of included shells. The results of our effective
 Êmulti-level recombination code were directly compared with the
 Êoutput from {\sc Recfast}.}
\label{fig:DN_eff_n}
\end{figure}

In Fig.~\ref{fig:DN_eff_n} we show the changes in the recombination
dynamics with the number of shells that were included into the
computation of the effective rates.
This figure confirms that our implementation of the effective
multi-level approach yields corrections that are in agreement with our
earlier computations \citet{Chluba2010}.
We find that the correction converges down to $z\sim 200$ when
including $\sim 300-400$ shells, as already explained in
\citet{Chluba2010}.
\changeI{We also directly compared with our full multi-level recombination code and found the difference to be smaller than $\Delta N_{\rm e}/N_{\rm e} \sim 10^{-5}$.}

Collisional processes are still able to change the low redshift
behaviour at the $\sim 0.1\%$ level in this redshift range
\citep{Chluba2010}, however, we defer a detailed analysis on the
importance of this effect to a future work.

\subsection{The reference case}
\label{sec:reference}
In Fig.~\ref{fig:DN_Hirata} we present a compilation of different
corrections to the ionization history that are included into our
reference case.
For this we {\it internally} compared the outputs of our recombination
code when switching on and off different processes.
We computed the solution to the photon transfer problem including the
Lyman-series up to $n=8$, with Lyman-$\theta$ ($n=9$) on the upper
boundary of the frequency grid.
In the Lyman-series transfer we did not include the corrections to the
{\it profiles} of the emission and absorption processes arising from
two-photon and Raman events, i.e. we described Lyman-$k$ emission and
absorption using Eq.~\eqref{eq:Line_em_ab_k}.
However, in our full reference case, Lyman-$k$ resonance and electron
scattering, as well as 2s-1s two-photon emission and absorption were
included (see Sect.~\ref{sec:DI-Ly-n_cases} for additional comments).

To account for all the corrections to the rate equations in the
effective multi-level recombination code, we ran the obtained solution
for the photon distribution through the modules that also allow us to
take the two-photon and Raman scattering corrections into account (see
explanations in Sect.~\ref{sec:DNnu_pert}). However, we replaced the
full profiles of the channels with the normal Voigt-profiles.

The cumulative difference with respect to the output of our effective
multi-level recombination code which does not include any of the
radiative transfer corrections is shown in Fig.~\ref{fig:DN_Hirata}.
In total we find a delay in recombination by $\Delta N_{\rm e}/N_{\rm
 Êe}\sim 0.4\%$ at $z\sim 930$, and an early acceleration by $\Delta
N_{\rm e}/N_{\rm e}\sim -1.0\%$ at $z\sim 1270$.
The reference case therefore captures a significant part of the total
correction with respect to {\sc Recfast} (see Sect.~\ref{sec:total_Ne}
for details).

\subsubsection{The 2s-1s two-photon correction}
\label{sec:DNe_2s-1s_2g}
\begin{figure}
\centering
\includegraphics[width=\columnwidth]{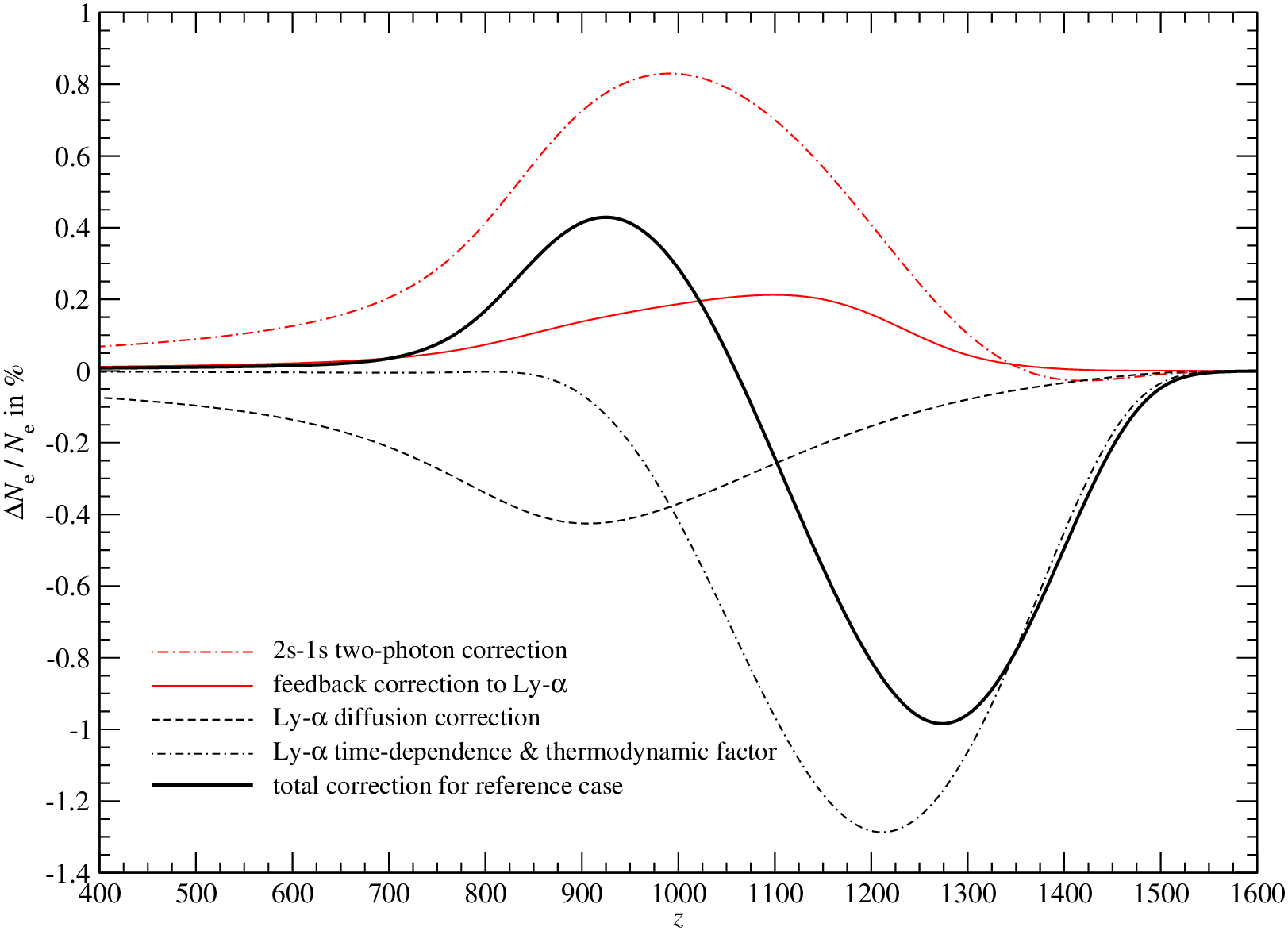}
\caption{Corrections that are included into the reference case.}
\label{fig:DN_Hirata}
\end{figure}

Figure~\ref{fig:DN_Hirata} shows the total correction due to changes
in the 2s-1s two-photon channel.
We only modified the 2s-1s two-photon and Ly-$\alpha$ net rate in our
effective multi-level recombination code using
Eq.~\eqref{eq:2s_1s_correct}, but did not alter any of the other
rates.
Also we switched off line-diffusion.

We find a delay of recombination by $\Delta N_{\rm e}/N_{\rm e}\sim
0.83\%$ at $z\sim 990$, which is slightly (by $\Delta N_{\rm e}/N_{\rm
 Êe}\sim 0.18\%$) larger than in earlier computations of this process
\citep[e.g.][]{Fendt2009}.

There are two main reasons for this difference; (i) because in the
reference case we include the emission and absorption in the 2s-1s
two-photon channel, the {\it self-feedback} of photons emitted by
2s-1s transitions on the 1s-2s two-photon channel is accounted for,
which leads to an additional delay of $\Delta N_{\rm e}/N_{\rm e}\sim
0.08\%$ and (ii) the remaining deceleration by $\Delta N_{\rm
 Êe}/N_{\rm e}\sim 0.1\%$ is just caused by normal absorption of 2s-1s
photons by Lyman-$\alpha$ (without the aid of line-diffusion).

\subsubsection{Feedback to Lyman-$\alpha$ and the diffusion correction}
\label{sec:DNe_Ly-a_feedback_diff}
We now considered the feedback correction to Ly-$\alpha$. Like in our
earlier treatment \citep{Chluba2007b} we find $\Delta N_{\rm e}/N_{\rm
 Êe}\sim 0.21\%$ at $z\sim 1100$.
We computed this correction from our radiative transfer code by
modifying the Lyman-$\alpha$ escape probability, with resonance
scattering switched off.
We also left the rate equations for the higher Lyman-series resonances
unaffected, in order to not reflect the full Lyman-series
feedback correction, which amounts to $\Delta N_{\rm e}/N_{\rm e}\sim
0.26\%$ at $z\sim 1100$ \citep{Chluba2009c}.

In Figure~\ref{fig:DN_Hirata} the correction due to Lyman-$\alpha$
diffusion alone is depicted. Again this was computed as a correction
to the Lyman-$\alpha$ resonance only.
We find an acceleration by $\Delta N_{\rm e}/N_{\rm e}\sim -0.44\%$ at
$z\sim 900$.
This is slightly smaller than in our earlier computation
\citep{Chluba2009b}. The reason is simply that there we only included
3 shells into our computation. However, when including more than
$5-10$ shells the diffusion correction becomes slightly smaller,
reducing to the curve presented in Fig.~\ref{fig:DN_Hirata}.
To check the precision of our PDE-solver, we recomputed the curve
for the 3-shells case and confirmed our earlier result.

\subsubsection{The correction due to time-dependence and thermodynamic factor}
\label{sec:DNe_thermo}
The final correction that is taken into account by the computations in
the reference case is caused by the time-dependence of the
Lyman-$\alpha$ emission process and the thermodynamic correction
factor (see Fig.~\ref{fig:DN_Hirata}). ÊThe origin of these terms \changeI{was
first explained in detail by} \citet{Chluba2008b} and \citet{Chluba2009}.
The net effect is an acceleration in recombination by $\Delta N_{\rm
 Êe}/N_{\rm e}\sim -1.28\%$ at $z\sim 1200$.
This result is in excellent agreement with the curves presented in
\citet{Chluba2009}, Fig.~18 therein.
Note that in the Figure of \citet{Chluba2009} also the 3s-1s and 3d-1s
two-photon profile correction was included.

\subsubsection{Correction from the higher Lyman-$n$}
\label{sec:DNe_Ly-n}
For the solid black line in Fig.~\ref{fig:DN_Hirata} all Lyman-series
corrections were included.
However, so far we have just discussed the corrections to the 2s-1s
two-photon and Lyman-$\alpha$ channel.
We found a cumulative acceleration by $\Delta N_{\rm e}/N_{\rm e}\sim
-0.06\%$ at $z\sim 1210$ as result of the higher Lyman-series.
This correction includes all feedback corrections among the higher
levels, time-dependence, and the thermodynamic factors.
{Since the additional modification is small, this} shows that just a 
detailed treatment of 2s-1s two-photon and
Lyman-$\alpha$ corrections already gives a very good approximation to
the total correction in the reference case.

\begin{figure}
\centering
\includegraphics[width=\columnwidth]{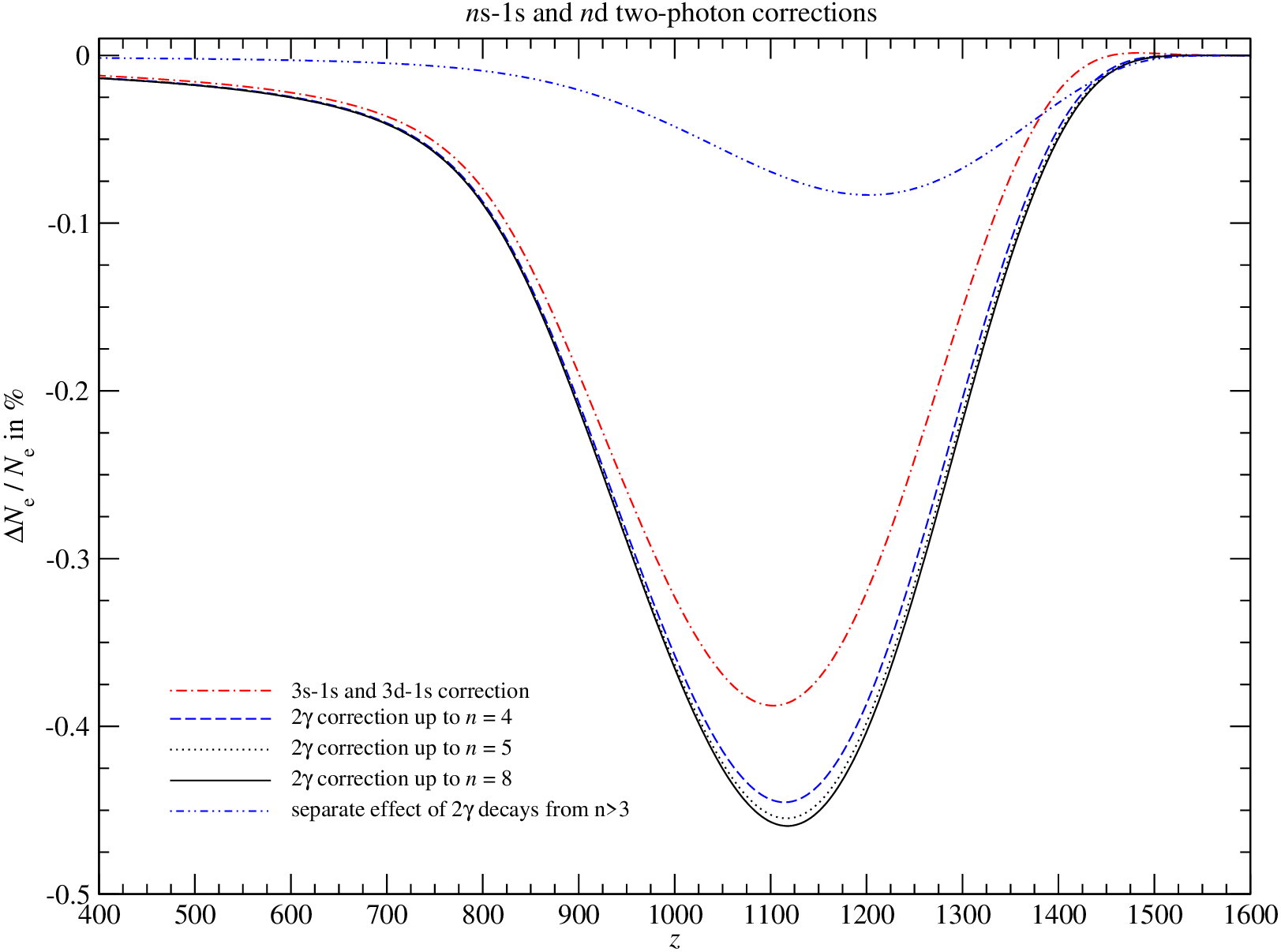}
\caption{Two-photon corrections from highly excited levels ($n>2$) and their convergence with $n$.}
\label{fig:DN_two-gamma}
\end{figure}
\subsection{Two-photon corrections from levels with $n>2$}
\label{sec:Two-gamma}
In this Section we discuss the correction caused by the modifications
in the emission and absorption profiles of the $n$s-1s and $n$d-1s
two-photon channels.
These corrections are due : (i) {\it quantum-mechanical} modifications
to the shapes of the line-profiles, and (ii) {\it stimulated}
two-photon emission in the CMB blackbody radiation field.
For two-photon processes from $n$s and $n$d-states with
$n>2$ the former dominates.

In Fig.~\ref{fig:DN_two-gamma} we present the changes to the free
electron fraction when the two-photon corrections up to the 8s-1s and
8d-1s two-photon channel are included. Two-photon processes lead to a
total acceleration of recombination by $\Delta N_{\rm e}/N_{\rm e}\sim
-0.46\%$ at $z\sim 1120$.
The main contribution comes from the 3s-1s and 3d-1s two-photon
process, while the higher levels only add $\Delta N_{\rm e}/N_{\rm
 Êe}\sim -0.08\%$ at $z\sim 1200$.
The correction practically converge when accounting for the
two-photon terms up to 5s-1s and 5d-1s.
Also the result for the 3s-1s and 3d-1s two-photon process compares
extremely well with our earlier computation \citep{Chluba2009}.
We conclude that for practical purposes it is sufficient to include
the two-photon corrections for all $n$s and $n$d states up to $n\sim
4-5$.

\begin{figure}
\centering
\includegraphics[width=\columnwidth]{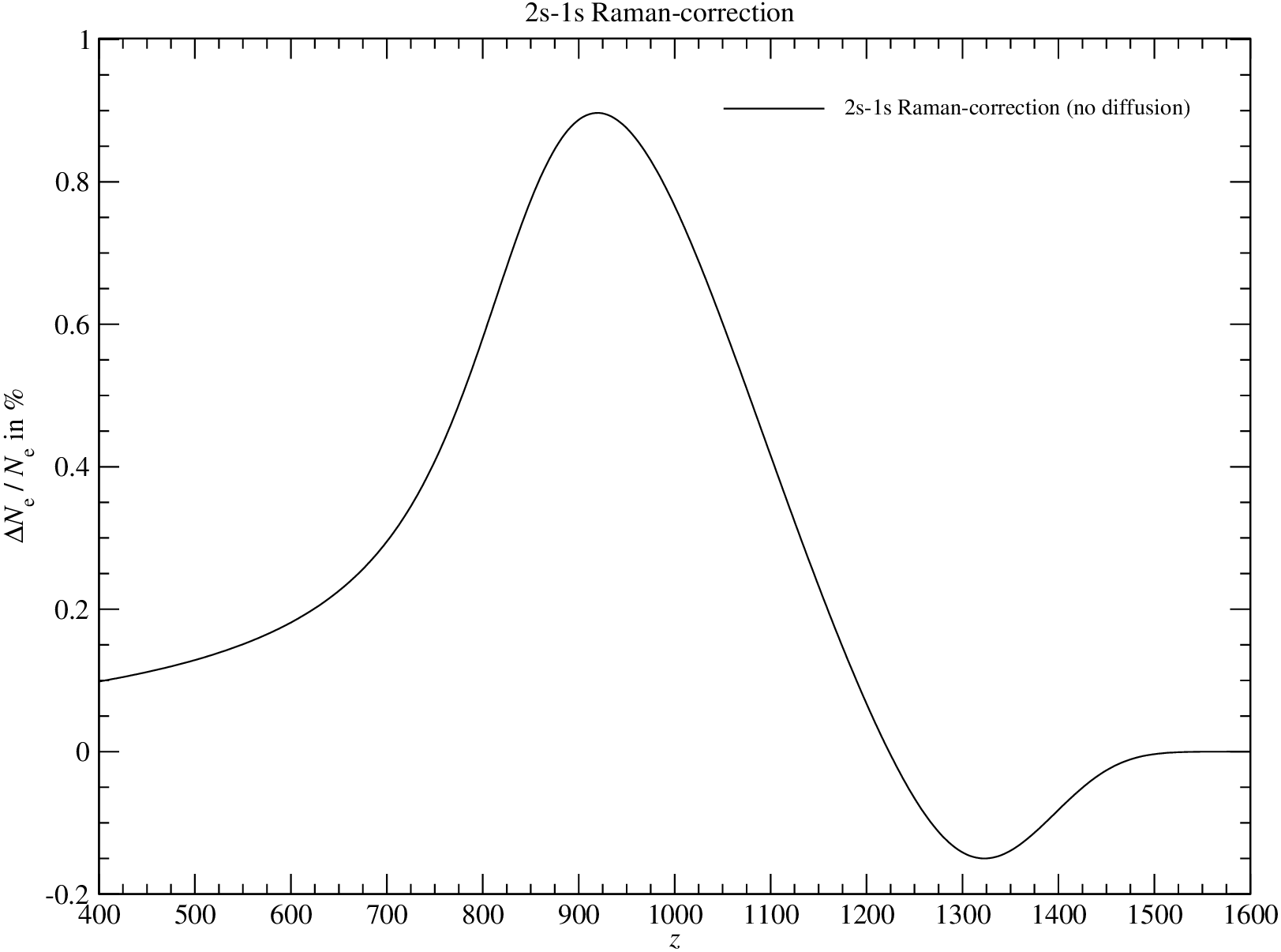}
\caption{Raman scattering corrections from excited levels}
\label{fig:DN_Raman}
\end{figure}
\subsection{Corrections caused by Raman processes}
\label{sec:Raman}
The final process we discuss is the effect of Raman scattering, which
was investigated also by \citet{Hirata2008}.
The result of our computation is shown in Fig.~\ref{fig:DN_Raman},
confirming that Raman scattering leads to a delay of recombination by
$\Delta N_{\rm e}/N_{\rm e}\sim 0.9\%$ at $z\sim 920$.
{This result is in very good agreement with the analysis of \citet{Hirata2008}.}
We found that the correction {is dominated} by the 2s-1s
Raman process. Higher level Raman scattering lead to a small
additional modification, which for practical purposes {could} be
neglected.
{We recommend including the Raman-corrections for the first three shells.}

As mentioned above, the Raman correction has two separate
contribution: one from the feedback of Lyman-$\beta$ photons on the
Lyman-$\alpha$ resonance, which leads to a delay of recombination a
low redshifts, and a second arising because of the accelerated 2s-1s
scattering. ÊThe delay and acceleration need to be out-of-phase in
redshift to create a net effect.
{We find that their individual contributions have amplitudes comparable to $\sim 2\%$, 
however, partial cancelation makes them smaller.}

\subsection{Total correction with respect to {\sc Recfast}}
\label{sec:total_Ne}
\begin{figure}
\centering
\includegraphics[width=\columnwidth]{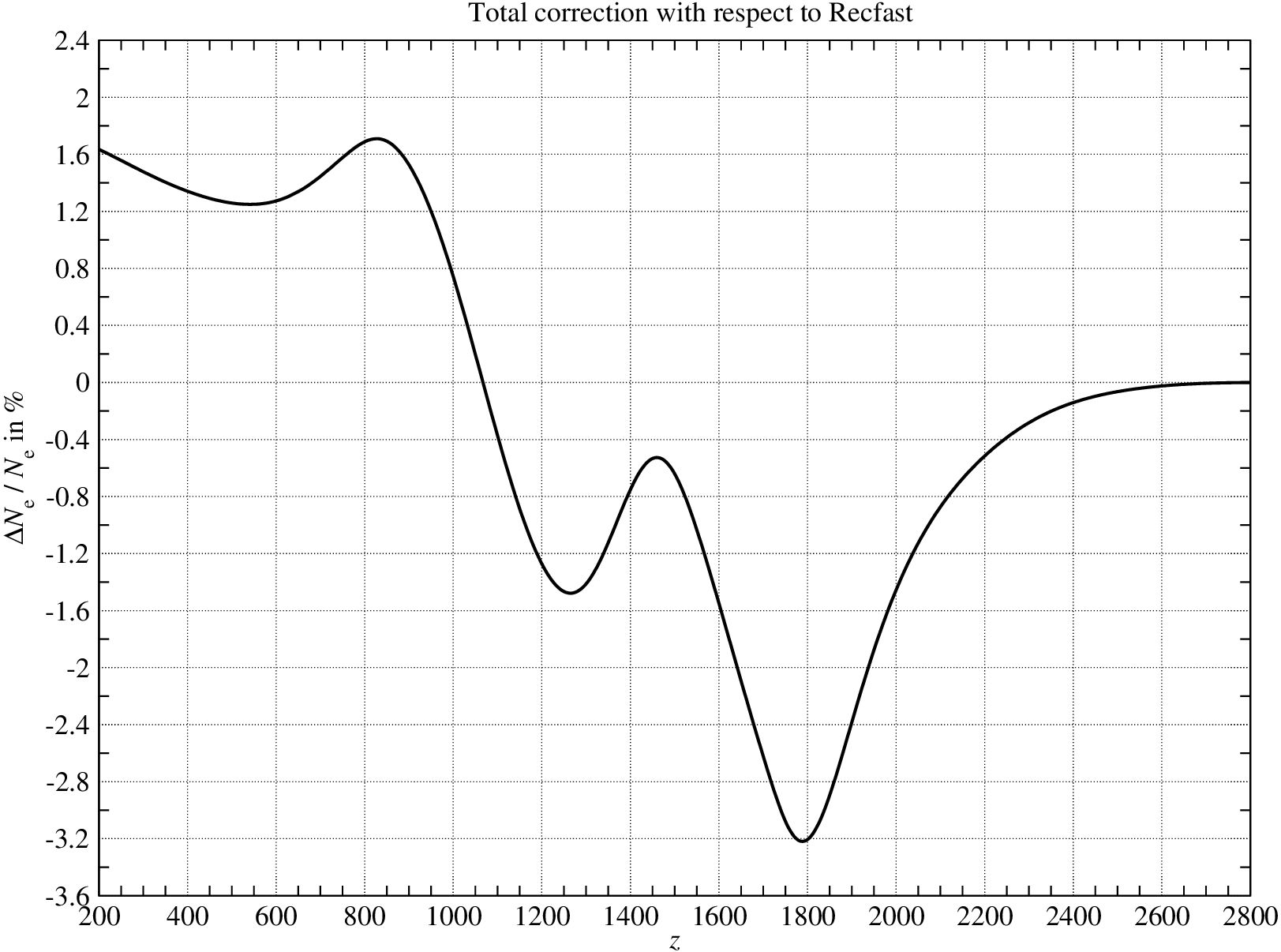}
\caption{Total correction to the ionization history. We compared the output of our effective multi-level recombination code with {\sc Recfast v1.4.2} \citep{Wong2007}. However, we switched all helium flags in {\sc Recfast} off.}
\label{fig:DN_total}
\end{figure}

In Fig.~\ref{fig:DN_total} we show the cumulative correction to the
ionization history caused by all the processes included into our
present recombination code.
The changes during hydrogen recombination found here are very similar
to those presented in \citet{Jose2010}.
The only major difference is visible at low redshifts, since their
analysis was based on the results from a 100-shell hydrogen
recombination model and hence the low redshift freeze-out tail was
overestimated (see Fig.~\ref{fig:DN_Hirata} and comments in
\citet{Chluba2010}).
However, as argued earlier \citep{Fendt2009, Jose2010, Chluba2010}, we
expect that this additional modification does not affect the
conclusions of their work at a significant level.
In particular, they showed that the cumulative effect of all published
recombination corrections could lead to biases in the values of
$\Omega_{\rm b}\,h^2$ and $n_{\rm S}$, that reach $\sim -1.7\sigma$
and $\sim -2.3\sigma$, respectively.
For the analysis of CMB data from the {\sc Planck} Surveyor these
corrections have to be taken into account carefully when answering 
queries about different models of {\it inflation}.

\section{Conclusions}
\label{sec:conc}
In this paper we complete our analysis on the importance of two-photon
transitions and Raman scattering during the cosmological recombination
epoch.
We explicitly solve the radiative transfer equation for the \ion{H}{i}
Lyman-series transport, including all important processes
(e.g. resonance scattering, full time-dependence), extending our
former treatment to account for Raman  {scattering, as well as two-photon transition from highly excited levels with $n>3$}.
Our computations are performed using an effective multi-level approach
for hydrogen to accelerate the recombination calculation, {that} without
optimization {achieves} runtimes of $\sim 1-2$ minutes.

We find that 2s-1s Raman scattering leads to a small initial
acceleration of recombination at high redshifts, which then turns into
a deceleration of $\Delta N_{\rm e}/N_{\rm e}\sim 0.9\%$ at $z\sim
920$.
$n$s-1s and $n$d-1s Raman processes from levels with $n>2$ only {result in} a small additional correction.
Two-photon transitions from $n$s and $n$d-states with $n>3$ accelerate
hydrogen recombination by additional $\Delta N_{\rm e}/N_{\rm e}\sim
-0.08\%$ at $z\sim 1200$. For practical purposes one only has to
include the two-photon corrections for the first $\sim 4-5$ shells.

This work carves a path towards a new cosmological
recombination code, \changeI{{\sc CosmoRec}}, that supersedes the physical model included in {\sc
  Recfast} and can be used in the analysis of future CMB data,
e.g. from the {\sc Planck} Surveyor, {\sc Act}, {\sc Spt}, and {\sc
  CmbPol}.
 The final step will be to perform a detailed code comparison, and to
optimize the implementation of the recombination code, so that
runtimes of seconds can be accomplished, incorporating all the
important physical processes {without} requiring any fudge factors.
\changeI{Our final version of {\sc CosmoRec} will be available at www.Chluba.de/CosmoRec.}

\section*{Acknowledgements}
\changeI{The authors thank the anonymous referee for comments on the 
manuscript, which helped improve the paper.}
JC wishes to thanks Chris Hirata, Yacine Ali-Haimoud and Dan Grin for
useful and stimulating discussion about recombination processes.
Furthermore, JC is very grateful for additional financial support from
the Beatrice~D.~Tremaine fellowship 2010.  Also, the authors
acknowledge the use of the GPC supercomputer at the SciNet HPC
Consortium. SciNet is funded by: the Canada Foundation for Innovation
under the auspices of Compute Canada; the Government of Ontario;
Ontario Research Fund - Research Excellence; and the University of
Toronto.

\begin{appendix}

\section{Computing the two-photon emission and Raman profiles}
\label{app:comp_two_photon_profs}

\subsection{Two-photon emission profiles}
We compute the two-photon decay profiles according to
\citet{Chluba2008b} and \citet{Chluba2009}.
In their treatment the infinite sum over intermediate $n$p-states is
split up into levels with principal quantum numbers $n>n_i$ and $n\leq
n_i$, where $n_i$ is the principal quantum number of the initial
state. This makes the sum over resonances (in the case of 3s and
3d only one) finite and {allows to use interpolation or fitting formulae 
for the remaining contributions} to the total matrix element from the infinite sum. 

We tabulate the non-resonant parts of the two-photon matrix elements
prior to the computation. The resonances are analytically added
afterwards.
As explained in \citet{Chluba2009}, close to the resonances, the
motion of the atoms becomes important, leading to a broadening of the
two-photon profiles. To include the effect of motion on the shape of
the lines close to the Doppler core, we take the ratio,
$\rho^{2\gamma}_i=\varphi^{2\gamma}_i/\varphi^{\Sigma\Lambda}_i$, of
the vacuum two-photon profiles to the sum of Lorentzians,
and tabulate it on the computational grid in frequency.
For every evaluation in time, we first compute the
Voigt-profiles for the resonances of interest and then sum these
with their respective weights to obtain the total Voigt-profile of
the resonance. These are then multiplied {by} $\rho^{2\gamma}_i$ to obtain an approximation for the two-photon
profile in the lab frame.

This procedure also allow us to include the changes in the total
width, $\Gamma_{n\rm p}$, of the resonances with redshift. In vacuum
this width is related to the total decay rate, however, with the CMB
this rate can change at the level of $\sim 10\%$ for the Lyman-$n$
line when $n>2$.
%

\begin{figure}
\centering
\includegraphics[width=\columnwidth]{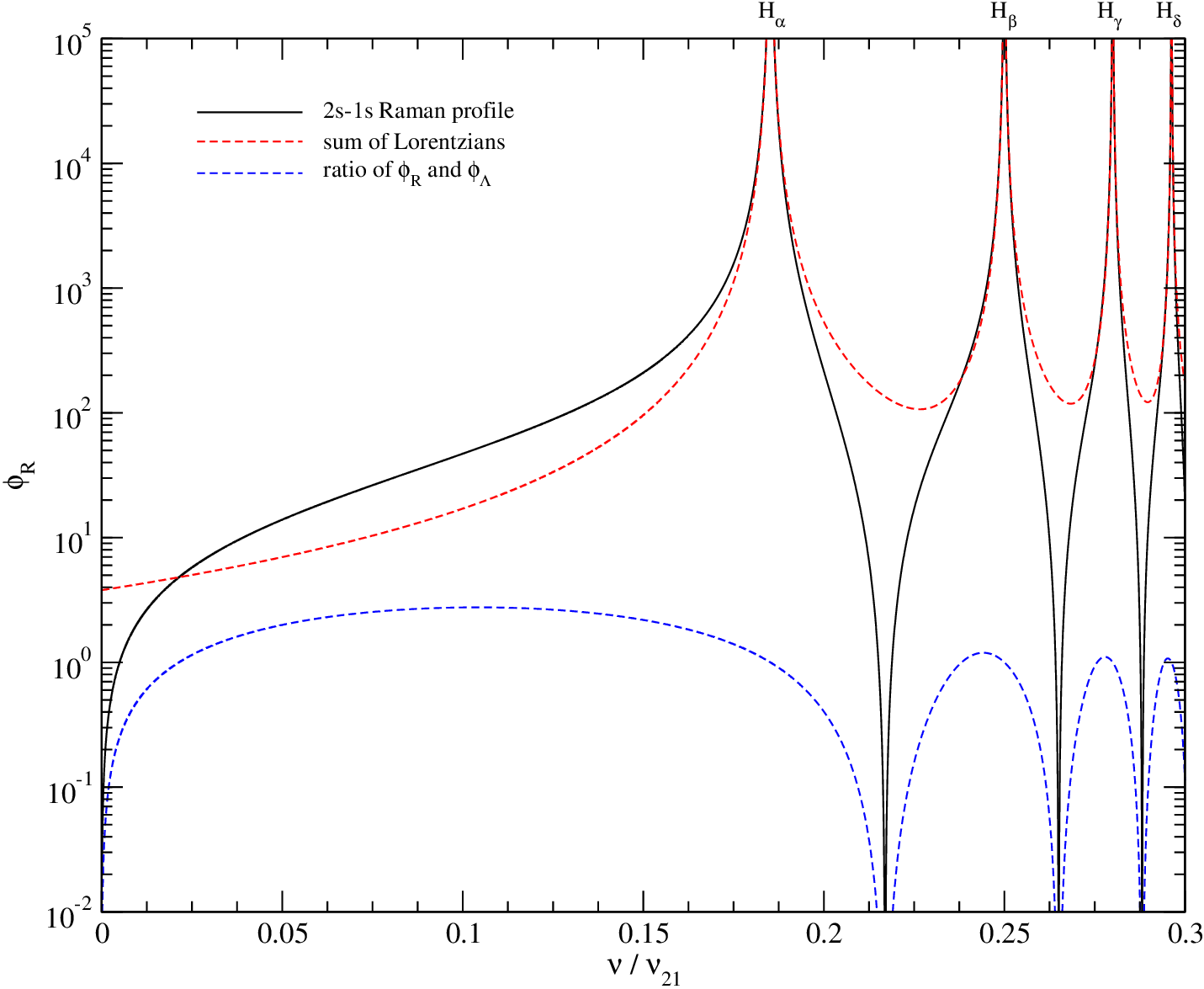}
\caption{Frequency dependence of the 2s-1s Raman profile. This profile
  has to be interpreted as a scattering cross section, where the
  electron is in the excited state. The positions of the
  Balmer-$\alpha$, $\beta$, $\gamma$, and $\delta$ resonances are
  marked. For comparison the cross section computed as a sum of
  Lorentzians are shown. Furthermore, we also show the ratio of these
  two profiles, indicating that close to the resonances the profile
  becomes Lorentzian, with small corrections.}
\label{fig:Phi_Raman}
\end{figure}

\begin{figure}
\centering
\includegraphics[width=\columnwidth]{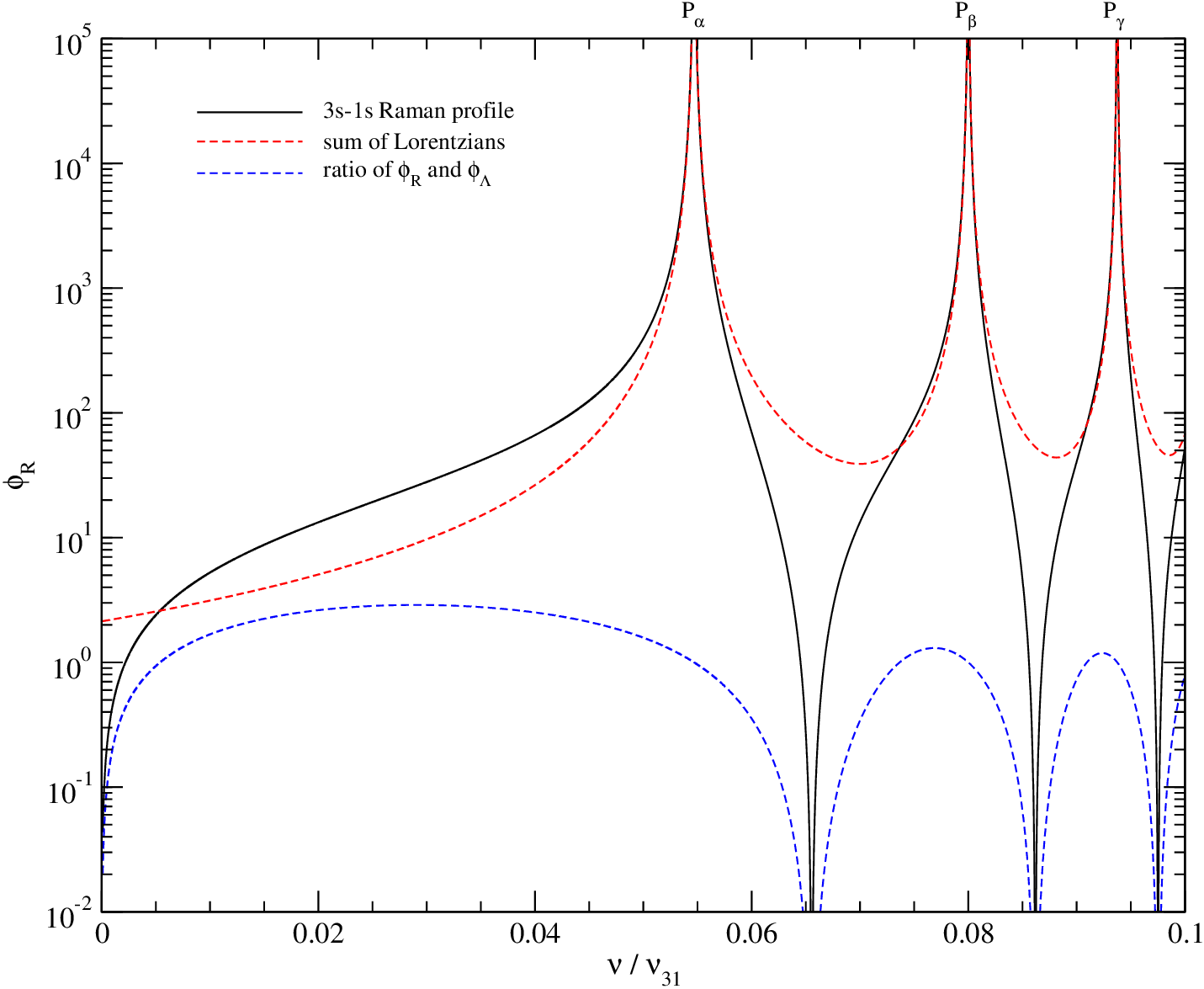}
\\[1mm]
\includegraphics[width=\columnwidth]{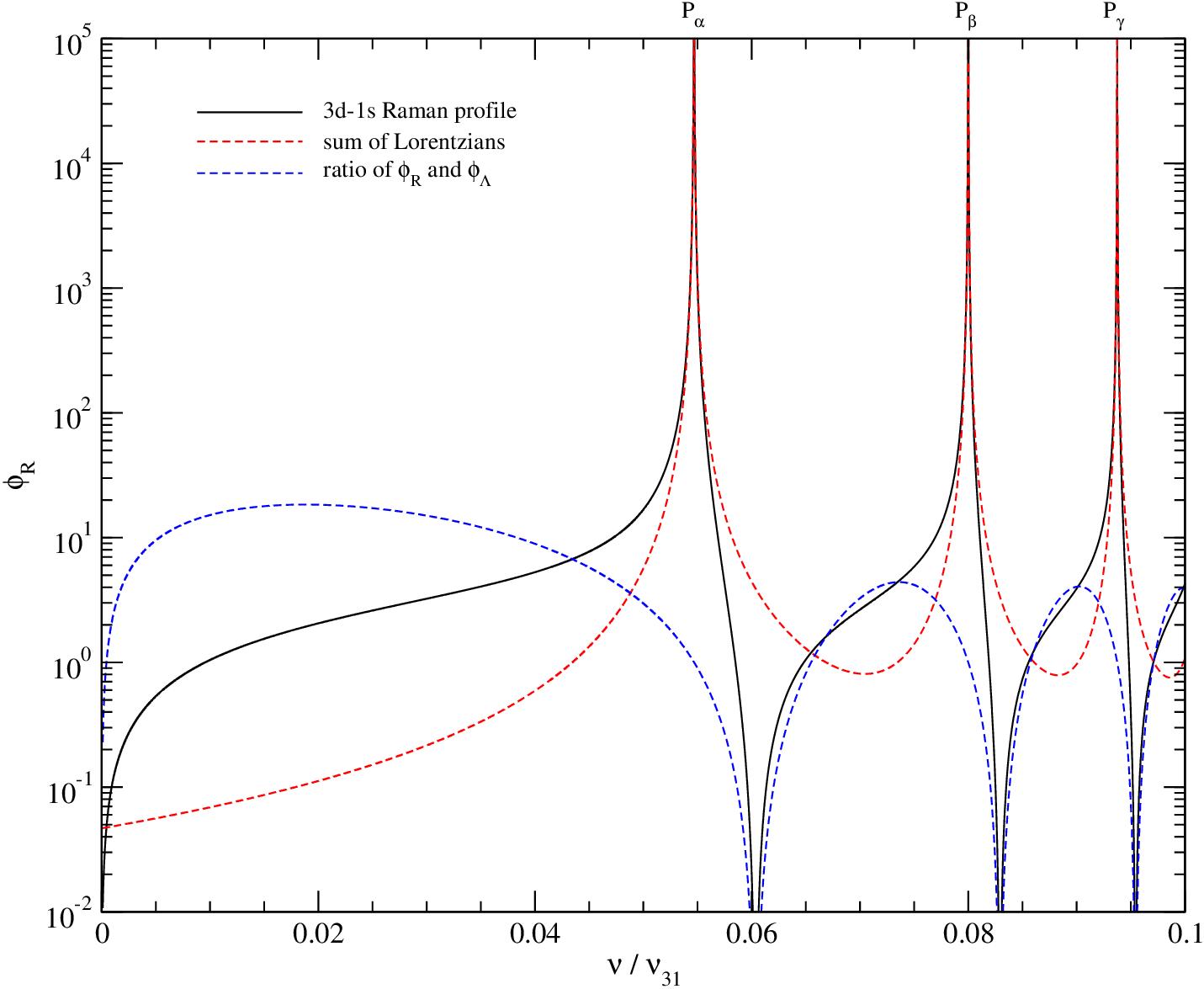}
\caption{Same as Fig.~\ref{fig:Phi_Raman}, but for the 3s-1s and 3d-1s Raman profiles. We marked the positions of the Paschen-$\alpha$, $\beta$, and $\gamma$ resonances.}
\label{fig:Phi_Raman.3s3d}
\end{figure}

\subsection{Raman profiles}
\label{app:Raman_Profiles}
A $n$s-1s and $n$d-1s Raman process has form ${\rm H} ^\ast +
\gamma\rightarrow {\rm H} + \gamma'$, where ${\rm H} ^\ast$ denotes a
neutral hydrogen atom in an excited $n$s/$n$d-state.
In contrast to the two-photon emission process, ${\rm H} ^\ast
\rightarrow {\rm H} + \gamma + \gamma'$, the Raman process only works
when photons are available. In vacuum no Raman scattering events occur.
The $n$s-1s and $n$d-1s Raman scattering matrix elements are related
to the $n$s-1s and $n$d-1s two-photon matrix elements by {\it
  crossing-symmetry}.
The energies of the incoming photon, $\gamma$, and the outgoing
photon, $\gamma'$, are given by $\nu+\nu_{j\rm 1s}=\nu'$, where
$h\nu_{j\rm 1s}$ is the excitation energy of the initial state with
respect to the ground state.

{
Raman scattering profiles can be derived using the formulae given in \citet{Chluba2009}. 
The main difference are: (i)  in the functions $f_n(y)$ and $h_n(y)$ (see Eq.~(8c) and (10d) in their paper) $y=\nu/\nu_{j\rm 1s}$ has to be replaced with $-y$; (ii) the pre-factor $y^3(1-y)^3$ needs to be substituted by $y^3(1+y)^3$; and (iii) the resonances now appear for intermediate $n$p-states with $n>n_j$, where $n_j$ is the principle
quantum number of the initial state.
}

For a given computational frequency grid only a finite number of
resonances appear, say for $n_j<n\leq n_{\rm res}$. Like in
computations of the two-photon emission profiles, one can therefore
split the infinite sum over intermediate p-states (including the
continuum), into resonant and non-resonant contributions. The
non-resonant contributions come from $n<n_j$ and $n_{\rm res}<n$,
where the non-resonant matrix element scales like $1/y$ for
$y\rightarrow 0$ and $(1/[n^2_j-1]-y)^{-1}$ towards the ionization
threshold, $\nu\rightarrow \nu_{j\rm c}$. One can therefore tabulate
$M_{\rm nr}\,y\,[1/[n^2_j-1]-y]$ on a grid and add the finite number
of remaining resonances analytically.
The pole displacements arising from the finite lifetime of the
intermediate p-state, as mentioned in \citet{Chluba2009}, has to be
included.

In Fig.~\ref{fig:Phi_Raman} and \ref{fig:Phi_Raman.3s3d} we present some examples of Raman scattering profiles. 
The electron is assumed to be in the excited state, so that a low frequency photon can Raman scatter off the atom.
In the recombination problem, these photons will be drawn from the CMB blackbody, as spectral distortions below the Balmer continuum can be neglected.
Figures~\ref{fig:Phi_Raman} and \ref{fig:Phi_Raman.3s3d}, also show the ratios of the Raman profiles with respect to the sum of Lorentzians. Close to the resonances all these ratios are extremely close to unity, as expected.
We use this ratio to include the effect of motion of the atom on the shape of the resonances close to the Doppler core, as explained in the previous section.

\section{PDE-solver}
\label{app:PDE_solver}

For this work we implement our own partial differential equation (PDE)
solver in order to fine tune performance and precision. Comparing
with previous results obtained using the NAG library confirms the
precision of our own implementation.

The PDE describing the radiative transfer problem during recombination
is of the {\it parabolic} type.
It is desirable to use an implicit or semi-implicit numerical scheme,
to avoid strong limitation on the step size imposed by stability.
Several numerical algorithms for this type of problems have been
discussed, e.g. Crank-Nicolson method \citep[see][]{Antia2002}.

For the recombination problem it is beneficial to use a {\it
non-uniform} grid in frequency as in the vicinity of the resonances
one needs a resolution of $\Delta\nu/\nu \sim 10^{-7}-10^{-5}$, {while} a much coarser grid can be introduced outside this
zone.
However, this implies that the spatial discretization that is normally
used in the Crank-Nicolson method is only accurate to first order in
the grid spacing. We therefore decided to implement a second order
scheme in which the first and second derivatives of the occupation
number $n_x$ with respect to frequency can be written as
\beal
\label{app:derivs}
\pAb{n_x}{x} &=\sum_i \kappa_i(x)\,n_{x_i}
\\
\PAb{n_x}{x}{2} &=\sum_i \Lambda_i(x)\,n_{x_i}
\end{align}
where the sums run over five grid-points in the neighbourhood of
$x$. The coefficients $\kappa_i(x)$ and $\Lambda_i(x)$ can be easily
derived using Lagrange interpolation formulae
\citep[e.g. see][]{Stegun1972}.
These coefficients can then be precomputed and stored once the grid is chosen.
The PDE appearing in the diffusion problem can thus be written as
matrix equation,
\beal
\label{app:Diffusion_Equ}
{\bf B_{ij}}\, n_{x_j} = b_{i},
\end{align}
where the matrix, ${\bf B_{ij}}$, is banded\footnote{At the boundary
  the matrix is not perfectly banded, but this does not
  pose a big problem.} with four off-diagonal elements.
Such system can be easily solved with $\mathcal{O}(M)$ operations,
where $M$ denotes the number of grid-points.

For each resonance we typically needed $\sim 10^3$ points in frequency. 
\changeI{Increasing this number to $\sim 10^4$ per resonance did not make a notable difference for the final correction to the ionization history.}
Our typical step size in redshift was $\Delta z \sim 1$, \changeI{but we also tried a ten times smaller time-step, without finding any significant modification in the solution. Our tests also showed that even $\Delta z\sim 10$ should be sufficient for detailed computations of the ionization history}. 

\changeI{Furthermore,} we tried fully implicit and semi-implicit schemes ($\theta$-method with $\theta
> 0.5$), \changeI{finding} good performance for $\theta\sim 0.6$.
We also experimented with the distribution of grid points, and found
that it is important to sample the Doppler core well.

\end{appendix}

\bibliographystyle{mn2e}
\bibliography{Lit}

\end{document}